\documentclass[pdflatex,sn-nature]{sn-jnl}

\usepackage{graphicx}%
\usepackage{multirow}%
\usepackage{amsmath,amssymb,amsfonts}%
\usepackage{amsthm}%
\usepackage{mathrsfs}%
\usepackage[title]{appendix}%
\usepackage{xcolor}%
\usepackage{textcomp}%
\usepackage{manyfoot}%
\usepackage{booktabs}%
\usepackage{algorithm}%
\usepackage{algorithmicx}%
\usepackage{algpseudocode}%
\usepackage{listings}%
\usepackage{siunitx}
\usepackage{accents}
\usepackage{setspace}
\usepackage{tabularx}

\theoremstyle{thmstyleone}%
\theoremstyle{thmstyletwo}%

\theoremstyle{thmstylethree}%

\usepackage[numbers]{natbib}
\renewcommand{\cite}[1]{\textsuperscript{\citealp{#1}}}

\newcommand\figref[1]{Fig.~\!\ref{fig:#1}}
\newcommand\figureref[1]{Figure~\ref{fig:#1}}

\newcommand\ee[0]{\mathrm{e}}
\newcommand\dd[0]{\mathrm{d}}

\newcommand{\ellipse}{\raisebox{-1pt}{\scalebox{1.3}[.4]{$\circ$}}}
\newcommand{\halo}{\accentset{\ellipse}}
\newcommand{\tp}{^{\top}}

\usepackage{ragged2e} 
\newcommand{\ex}[1]{\langle#1\rangle}
\newcommand{\xt}{\ex{\hat{\bf x}}_{\rm T}}
\newcommand{\xf}{\ex{\hat{\bf x}}_{\rm F}}
\newcommand{\xr}{\ex{\hat{\bf x}}_{\rm R}}
\newcommand{\xs}{\ex{\hat{\bf x}}_{\rm S}}
\newcommand{\erf}[1]{Eq.~\eqref{}}

\newcommand\gammaTc[0]{\gamma^{}_{\hbox{\scalebox{0.5}{$T_c$}}}}
\newcommand\gammaLc[0]{\gamma^{}_{\hbox{\scalebox{0.5}{$L_c$}}}}

\newcommand{\fil}{_{\rm F}}
\newcommand{\rfil}{_{\rm R}}
\newcommand{\sm}{_{\rm S}}
\newcommand{\god}{_{\rm T}}
\newcommand{\inv}{^{-1}}
\newcommand{\bx}{{\bf x}}
\newcommand{\bz}{{\bf z}}
\newcommand{\dg}{^\dagger}
\newcommand{\beq}{\begin{equation}}
\newcommand{\eeq}{\end{equation}}

\newcommand{\Tr}{\text{Tr}}
\usepackage{color}

\usepackage[normalem]{ulem}

\definecolor{nbrown}{rgb}{0.5,0.4,0.0}

\definecolor{npurple}{rgb}{0.627, 0.125, 0.941}

\begin{document}

\title[Article Title]{Tracking Quantum Dynamics in an Optical Cavity for Recovering Purity and Squeezing via Quantum State Smoothing}


\author[1,2,3]{\fnm{Shota} \sur{Yokoyama}}
\author[3,4,5,6]{\fnm{Kiarn T.} \sur{Laverick}}
\author[2,3]{\fnm{David} \sur{McManus}}
\author[2,3,4]{\fnm{Qi} \sur{Yu}}
\author[3,4,7]{\fnm{Areeya} \sur{Chantasri}}
\author[1,8]{\fnm{Warit} \sur{Asavanant}}
\author[2,3,9]{\fnm{Daoyi} \sur{Dong}}
\author[3,4]{\fnm{Howard M.} \sur{Wiseman}}
\author*[1,2,3]{\fnm{Hidehiro} \sur{Yonezawa}}\email{h.yonezawa@riken.jp}

\affil[1]{\orgname{RIKEN Center for Quantum Computing}, \orgaddress{\street{2-1 Hirosawa}, \city{Wako}, \state{Saitama} \postcode{351-0198}, \country{Japan}}}

\affil[2]{\orgdiv{School of Engineering and Technology}, \orgname{The University of New South Wales}, \orgaddress{\city{Canberra}, \state{ACT} \postcode{2600}, \country{Australia}}}

\affil[3]{\orgdiv{Centre for Quantum Computation and Communication Technology}, \orgname{Australian Research Council}}

\affil[4]{\orgdiv{Centre for Quantum Dynamics}, \orgname{Griffith University}, \orgaddress{\street{Yuggera Country}, \city{Brisbane}, \state{Queensland} \postcode{4111}, \country{Australia}}}

\affil[5]{\orgdiv{MajuLab}, \orgname{CNRS-UCA-SU-NUS-NTU International Joint Research Laboratory}}

\affil[6]{\orgdiv{Centre for Quantum Technologies}, \orgname{National University of Singapore}, \orgaddress{\postcode{117543}, \country{Singapore}}}

\affil[7]{\orgdiv{Optical and Quantum Physics Laboratory, Department of Physics, Faculty of Science}, \orgname{Mahidol University}, \orgaddress{\city{Bangkok}, \postcode{10400}, \country{Thailand}}}

\affil[8]{\orgdiv{Department of Applied Physics, School of Engineering}, \orgname{The University of Tokyo}, \orgaddress{\street{7-3-1 Hongo, Bunkyo-ku}, \city{Tokyo}, \postcode{113-8656}, \country{Japan}}}

\affil[9]{\orgdiv{Australian Artificial Intelligence Institute}, \orgname{University of Technology Sydney}, \orgaddress{\state{NSW} \postcode{2007}, \country{Australia}}}

\maketitle

\textbf{\noindent
Tracking the dynamics of a quantum system is conventionally achieved by monitoring the system continuously in time and filtering the information contained in measurement records via the \textit{causal} quantum trajectory approach. However, in practical scenarios there is often loss of information to the environment, leading to filtered states that are impure because of decoherence. If real-time tracking is not required, the lost information can be maximally extracted via \textit{acausal quantum state smoothing}, which has been theoretically proven to better restore the system's coherence (purity) than causal filtering. Interestingly, quantum state smoothing requires assumptions of how any lost quantum information (unobserved by the experimenter) was turned into classical information by the environment. In this work, we experimentally demonstrate smoothing scenarios, using an optical parametric oscillator and introducing `observed' and `unobserved' channels by splitting the output beam into two independent homodyne detectors. We achieve improvement in state purification of $\mathbf{10.3\% \pm 1.6\%}$, squeezing restoration of $\mathbf{7.6\% \pm 2.6\%}$, and show that smoothed states are better estimates of hidden true states than those from conventional filtering. The estimation techniques used in this paper are promising for many applications in quantum information that incorporate post-processing.
}

\begin{figure}[b!]\centering
\includegraphics[scale=1]{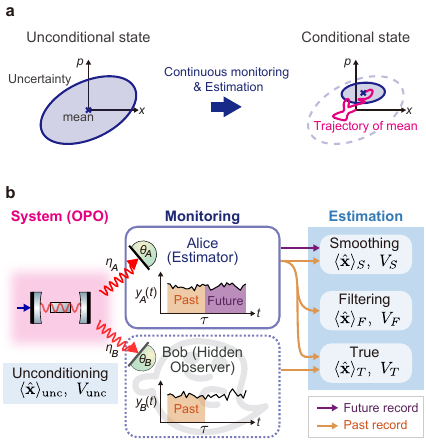}
\caption{\textbf{Abstract illustration for estimation of quantum dynamics.}
	\textbf{a}, Ellipses with cross marks and shaded areas in phase space ($x$ and $p$ coordinates) indicate the mean values and the quantum uncertainties of quantum states, respectively. The left figure shows an unconditional state, where the environment is not monitored, indicated by a state with large uncertainty due to decoherence. The right figure shows a conditional state, when the environment is monitored, with a trajectory of the mean in phase space (magenta curve) and a small quantum uncertainty (ellipse with shaded area).  
	\textbf{b}, An open quantum system, OPO (optical parametric oscillator), interacts with the environment involving monitoring (homodyne detection) devices with the measurement bases, $\theta_{\rm A}$ and $\theta_{\rm B}$, and the efficiencies $\eta_{\rm A}$ and $\eta_{\rm B}$. The quantum state dynamics are estimated using Alice's (observed) record, $\mathbf{y}_{\rm A}(t)$. However, for quantum state smoothing, we need to assume a hidden observer, Bob, whose measurement record, $\mathbf{y}_{\rm B}(t)$, is unavailable to Alice (see more details in text).
}
\label{fig:1}
\end{figure}

The dynamics of a quantum system that interacts with an environment is of great interest in current quantum science and technology research.
As the quantum system becomes entangled with an unobserved  environment, the system's state will decohere, becoming a statistical mixture of possible states. 
Decoherence is a considerable obstacle to realizing real-world applications of quantum information technology\cite{nielsen2010quantum,wiseman2009quantum}. However, if the environment can be observed or measured, even only partially, one expects the purity and the coherence of the quantum system to be recovered. There has been intensive research in this area of preserving coherence, and the most effective way is to continuously monitor the environment to track the dynamics of the quantum system\cite{murch2013observing,rossi2019observing}.

\figureref{1}{a} illustrates how estimation of quantum dynamics, conditioned on the environment's monitoring results, purifies the state.
This shows quantum dynamics for states represented by Gaussians in phase space. A quantum state before measurement and estimation, known as the unconditional state, has large uncertainty (\figref{1}{a} left). Continuous monitoring and estimation of the quantum state provides the stochastic trajectory of the mean value, resulting in the reduced uncertainty, \textit{i.e.}, a conditional state (\figref{1}{a} right).
To date, estimation of quantum dynamics has been demonstrated in several systems, including superconducting qubits\cite{murch2013observing,weber2014mapping,chantasri2016quantum,naghiloo2017quantum,foroozani2016correlations,tan2017homodyne} and mechanical resonators coupled to an optical cavity\cite{wieczorek2015optimal, rossi2019observing, meng2020mechanical}.
In these demonstrations, measurement records up to time $\tau$ were processed to estimate a quantum state at time $\tau$.
Such an approach, called quantum state filtering (or the conventional quantum trajectory)\cite{carmichael1993open,belavkin1999measurement,doherty1999feedback,wiseman2009quantum}, allows an estimation in real time, which is required for feedback control\cite{zhang2017quantum}.

Real-time estimation, however, is not always required. 
In situations where there is still information missing from the past measurement record and post-processing is allowed, a better estimate may be achieved by employing measurement records both before (past) and after (future) time $\tau$. 
This estimation technique that utilizes past-future information is known as smoothing in classical estimation theory\cite{haykin2001kalman,weinert2001fixed,van2013detection,brown2012introduction,einicke2012smoothing,friedland2012control}. 
Smoothing is a powerful technique to estimate time-varying parameters, and has been proved to be effective for estimating classical parameters in quantum systems\cite{wheatley2010adaptive,yonezawa2012quantum,iwasawa2013quantum}.
A smoothing technique has been proposed particularly for quantum state estimation, conditioned on the past-future measurement record\cite{guevara2015quantum}, showing that smoothing does give better estimates of quantum states with higher purity than that from the conventional filtering approach. The quantum state smoothing has recently been applied to general linear Gaussian systems\cite{laverick2019quantum,laverick2021linear}, which is experimentally demonstrated for the first time in this work. In particular, we realize quantum state smoothing, in an empirically verifiable way, in an optical parametric oscillator (OPO). This system is a critical resource that is widely used in quantum information technologies\cite{yonezawa2012quantum,larsen2021deterministic,wheatley2010adaptive,iwasawa2013quantum,asavanant2021time,roy2021spectral,aasi2013enhanced,walk2016experimental,chen2014experimental}.
	
One distinctive characteristic of quantum state smoothing is that, contrary to classical smoothing and to quantum state filtering, it requires additional assumptions about the missing information in the environment to be recovered. Specifically, it is necessary to model how the quantum information in an unobserved environment could have been turned into classical information about the system. In other words, how a hidden measurement might have occurred. The advantage of our system is that our losses and other forms of decoherence are so low that we can collect almost all the quantum information lost by the system, and split it into two parts. One part is observed by the party doing the smoothing, Alice, and the other part is observed by a hidden party, Bob. In our specific experiment on the OPO system, as illustrated in \figref{1}{b}, the splitting is done by a beam-splitter and both Alice's detector (denoted with a subscript `A'), and Bob's detector (with subscript `B') are homodyne detectors.

To simplify the discussion, say Bob has access to Alice's records, as well as his own, so that he is also playing the role of an omniscient observer. 
Given this set-up, one can see that Bob's state should be the purest state, which we call the `True' state denoted by $\rho_{\rm T}$. Because the OPO system with homodyne measurements is a linear Gaussian one\cite{laverick2019quantum,laverick2021linear}, this state is described by a Gaussian Wigner function\cite{weedbrook2012gaussian} with mean $\langle \hat{\bf x} \rangle_{\rm T}$ and covariance matrix $V_{\rm T}$, as shown in \figref{1}{b}. Then Alice, with only her observed record, but through the knowledge of Bob's measurement setting, can use quantum state smoothing with her past (orange) and future (purple) measurement records to obtain her smoothed state $\rho_{\rm S}$ (described by $\langle \hat{\bf x} \rangle_{\rm S}$ and $V_{\rm S}$). Since Alice's smoothed state uses more information in estimating the true state (unknown to Alice) than the filtered state $\rho_{\rm F}$ (described by $\langle \hat{\bf x} \rangle_{\rm F}$ and $V_{\rm F}$), it is a purer state, a more squeezed state, and \textit{most importantly} a better estimate of the true state of the OPO system, than is the filtered state. The amounts of purity and squeezing that are recovered, as well as the quality of the estimation, are dependent on both the measurement efficiency and the measurement settings (homodyne angles) chosen by both Alice \textit{and} Bob.

\subsection*{Results}
\addcontentsline{toc}{subsection}{Results}

\subsubsection*{Experimental setup}
\addcontentsline{toc}{subsubsection}{Experimental setup}

In our experiment, we employ a continuous-wave fiber laser at \SI{1550}{\nano\meter} as the main laser source.
The OPO, the target system, is a bow-tie cavity with a round-trip length of \SI{489}{\milli\meter}. 
The energy transmittance of the output coupler is 10\% for the fundamental wavelength of \SI{1550}{\nano\meter}. The OPO contains a periodically poled potassium-titanyl-phosphate (PPKTP) crystal as a nonlinear medium, and is pumped by the frequency-doubled beam at \SI{775}{\nano \meter} which is not resonant in the OPO. 
The interaction Hamiltonian is given by 
$
	\mathcal{H} \propto \xi [(\hat a^\dagger)^2 + \hat{a}^2],
$
where  $\xi \in [0, 1)$ is a normalized pump amplitude\cite{collett1984squeezing}, $\hat a $ and $\hat a^\dagger$ are annihilation and creation operators, with $\dagger$ denoting the adjoint of an operator. The corresponding canonical position and momentum operators are defined as $\hat{x} = (\hat a + \hat a^\dagger)\sqrt{\hbar/2}$ and $\hat p = -i(\hat a - \hat a^\dagger)\sqrt{\hbar/2}$ with commutation relation $[\hat x, \hat p]=i \hbar$ ($\hbar$ is the reduced Planck constant). The interaction Hamiltonian generates squeezing along the $p$ quadrature dependent on the pump amplitude $\xi$ with corresponding anti-squeezing along the conjugate $x$ quadrature. In the limit that the OPO is pumped at threshold ($\xi\to 1$), the squeezing in the $p$ quadrature reaches half the fluctuations of the vacuum state, with infinite anti-squeezing in the $x$ quadrature. We operate at $\xi = 0.70$, which is sufficient to obtain unconditional squeezing well below vacuum fluctuations.

The output beam from the OPO is first split by a variable beam splitter (BS) to divide the measurement records into two channels, for Alice and Bob, where the splitting ratio is adjustable by tuning the energy transmittance, $T$, of the BS. The transmitted beam of the BS is measured by a homodyne detector (Alice) with a relative phase $\theta_{\rm A}$ between a local oscillator and the beam, \textit{i.e.}, the quadrature $\hat x \cos \theta_{\rm A} +\hat p \sin \theta_{\rm A}$ is measured by Alice. Similarly, the reflected beam is measured by another homodyne detector (Bob) with a relative phase $\theta_{\rm B}$, \textit{i.e.}, the quadrature $\hat x \cos \theta_{\rm B} +\hat p \sin \theta_{\rm B}$ is measured by Bob. The homodyne signals are recorded by oscilloscopes, yielding time-varying signals $y_{\rm A}(t)$ and $y_{\rm B}(t)$ for Alice and Bob, respectively. 

The measurement efficiencies for Alice's and Bob's measurements can be determined directly from the transmittance $T$ of the BS.
Due to experimental imperfections such as optical loss, the efficiencies are calculated from $\eta_{\rm A} = 0.865T$ and $\eta_{\rm B} = 0.863(1-T)$, which gives the total efficiency $\eta_{\rm tot} = \eta_{\rm A} + \eta_{\rm B} \approx 0.86$. We note that, even though the total efficiency $\eta_{\rm tot} \approx 0.86$ is not exactly one, as for the ideal quantum state smoothing theory, it is high enough for the state conditioned on both Alice's and Bob's records to be close to pure (as the experimental data will show) and hence we are justified in using the term `true' for these states.
More details of the experimental setup and data processing can be found in the Methods and Supplementary Information.

As stated above, since the OPO with homodyne detection is a linear Gaussian quantum system\cite{weedbrook2012gaussian}, all states (including the smoothed quantum state\cite{laverick2019quantum,laverick2021linear}) of the relevant intracavity optical mode are fully described by the mean vector
	$\langle \hat{\bf x} \rangle$ 
and the covariance matrix 
	$V ={\rm Cov} (\hat{\mathbf{x}}, \hat{\mathbf{x}})\,, $
 where 
	$\hat{\mathbf{x}} \equiv \left( \hat x \,\,\, \hat p\right)\tp$ 
 and 	
 	${\rm Cov}(\hat{\mathbf{o}}_1,\hat{\mathbf{o}}_2) \equiv \frac{1}{2} \left\langle \hat{\mathbf{o}}_1^{} \hat{\mathbf{o}}_2\tp + (\hat{\mathbf{o}}_2^{}  \hat{\mathbf{o}}_1\tp)\tp\right\rangle - \left\langle  \hat{\mathbf{o}}_1 \right\rangle\left\langle  \hat{\mathbf{o}}_2 \right\rangle\tp$.
Here, $\langle \hat{O} \rangle$ denotes the expectation value of the operator $\hat{O}$ and $\top$ denotes the transpose.
 Moreover, the steady-state covariance matrices for all linear Gaussian states are time-independent functions of the dynamical and measurement parameters of the system, independent of the particular measurement records observed by Alice and/or Bob. As such, the estimation of the state reduces to determining the trajectory of the mean vector. 
In our experiment, we compare the performance of Alice's quantum state filtering and smoothing results, making use in this comparison of the true state (which is the filtered state Bob calculates using Alice's record in addition to his own).  
While quantum state filtering utilizes only the past portion of Alice's measurement records, quantum state smoothing utilizes Alice's past-future measurement records with knowledge only of Bob's homodyne angle. 
See Methods for the details of the estimation procedures.

\subsubsection*{Estimation of quantum dynamics by quantum state smoothing}
\addcontentsline{toc}{subsubsection}{Estimation of quantum dynamics by quantum state smoothing}

\begin{figure*}[b!]\centering
\includegraphics[scale=1]{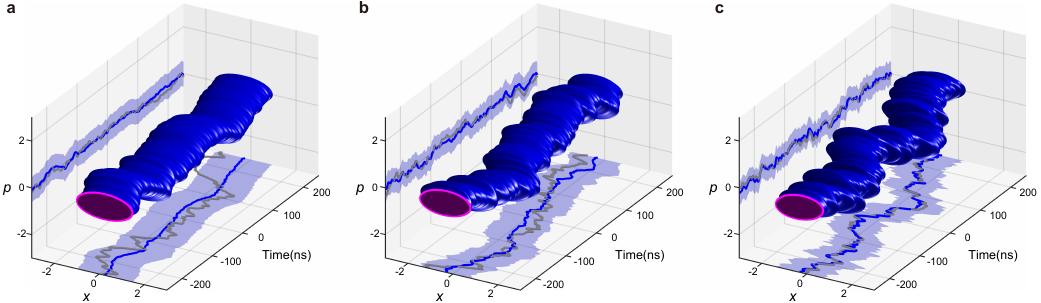}
\caption{\textbf{Estimated dynamics of the quantum state in an optical cavity by quantum state smoothing.}
	The means $\langle \hat{\bf x} \rangle_{\rm S}$ of the quadratures (blue curves) with their corresponding quantum uncertainties (one standard deviation shown as light blue regions) shown on the respective axes. The dark blue tube centered at $\langle \hat{\bf x} \rangle_{\rm S}$ is the $e^{-1/2}$ contour of the Wigner function (shown as a magenta ellipse at the earliest time).
	The grey traces show the trajectories of the true means, $\langle \hat{\bf x} \rangle_{\rm T}$, for comparison.
	The measurement efficiencies are different in three panels, with $\eta_{\rm A} = 0.09$ in \textbf{a}, 0.43 in \textbf{b}, and 0.78 in \textbf{c}.
	The measurement angles are set to $\theta_{\rm A} = \SI{65}{\degree}$ and $\theta_{\rm B} = \SI{135}{\degree}$ and 
        $\hbar$ is normalized to 1.
        }
\label{fig:Result1}
\end{figure*}

We first focus on the estimated dynamics of the state by quantum state smoothing.
\figureref{Result1} displays the estimated dynamics of the $x$ and $p$ quadratures with differing measurement efficiencies; $\eta_{\rm A} = 0.09$, $0.43$ and $0.78$ (\textbf{a}, \textbf{b} and \textbf{c} in \figref{Result1}, respectively).
The measurement angles are set to $\theta_{\rm A} = \SI{65}{\degree}$ and $\theta_{\rm B} = \SI{135}{\degree}$. 
The magenta ellipse in the cross section indicates the one standard deviation point of the Gaussian state calculated from the estimated smoothed variance $V_{\rm S}(t)$, centered about the smoothed mean $\ex{\hat{\bf x}}_{\rm S}$. 
In each panel of \figref{Result1}, the mean and one standard deviation of each quadrature is projected on the bottom and side walls. 
The true mean $\ex{\hat{\bf x}}_{\rm T}$ is also projected (grey trace) as an indicator for how the smoothed estimate is performing.
Immediately, we notice that as Alice's measurement efficiency is increased, the smoothed mean better approximates the true mean. This is to be expected as for low measurement efficiencies on Alice's side, Alice's measurement record only contains a small amount of information about the system. As the efficiency increases, Alice's measurement record contains more information about the true mean and the quality of estimation increase, as one would expect. More importantly, we can see that for each value of $\eta_{\rm A}$, the smoothed state mean is generally closer to the true state mean over the time interval than is the unconditioned mean $\ex{\hat{\bf x}}_{\rm unc} = 0$. These are all, however, qualitative statements of how well the smoothed estimate performs.

\begin{figure}[b!]\centering
\includegraphics[scale=1]{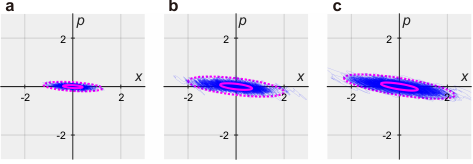}
\caption{\justifying\textbf{Trajectory traces of the smoothed means of $\hat x$ and $\hat p$ quadratures over the period of \SI{50}{\micro\second}.}
Traces of the smoothed-state mean trajectories (blue) are plotted in phase space.
The theoretical distribution for these means is a Gaussian with mean zero and covariance matrix $V_{\rm unc}-V_{\rm S}$. 
The magenta solid (dashed) ellipses show contours of this distribution at $e^{-1/2}$ ($e^{-9/2}$) of the peak value, corresponding to one (three) standard deviations for the variables aligned with the principal axes of these ellipses.
The experimental results show good agreement with the theory. The measurement efficiencies are changed from $\eta_{\rm A} = 0.09$ in \textbf{a}, 0.43 in \textbf{b}, and 0.78 in \textbf{c}. The measurement angles are set to $\theta_{\rm A} = \SI{65}{\degree}$ and $\theta_{\rm B} = \SI{135}{\degree}$ and $\hbar$ is normalized to 1. 
}
\label{fig:Result1-2}
\end{figure}

Since our system is Gaussian, the covariance of the unconditional state, $V_{\rm unc}$, is given by $V_{\rm S} + {\rm Cov}(\ex{\hat{\bf x}}_{\rm unc},\ex{\hat{\bf x}}_{\rm S})$, which equals $V_{\rm S} + {\rm Cov}(\ex{\hat{\bf x}}_{\rm S},\mathbf{0})$, since $\ex{\hat{\bf x}}_{\rm unc} = \mathbf{0}$. 
Thus, the distribution of the mean $\ex{\hat{\bf x}}_{\rm S}$ over time is a way to measure the reduction of the state variance, $V_{\rm unc}-V_{\rm S}$. 
The magenta solid and dashed circles in \figref{Result1-2} show respectively the $e^{-1/2}$ and $e^{-9/2}$ contours of this theoretical distribution, a zero mean bivariate Gaussian with covariance $V_{\rm unc}-V_{\rm S}$.
These contours do well represent the distribution of the trajectories in the figure. From this, we see that the larger the stochastic displacement in the smoothed mean, the purer the smoothed state, and ultimately the closer it is to the true state on average.

\subsubsection*{State purification and squeezing restoration via quantum estimation}
\addcontentsline{toc}{subsubsection}{State purification and squeezing restoration}

\begin{figure}[b!]\centering
\includegraphics[scale=1]{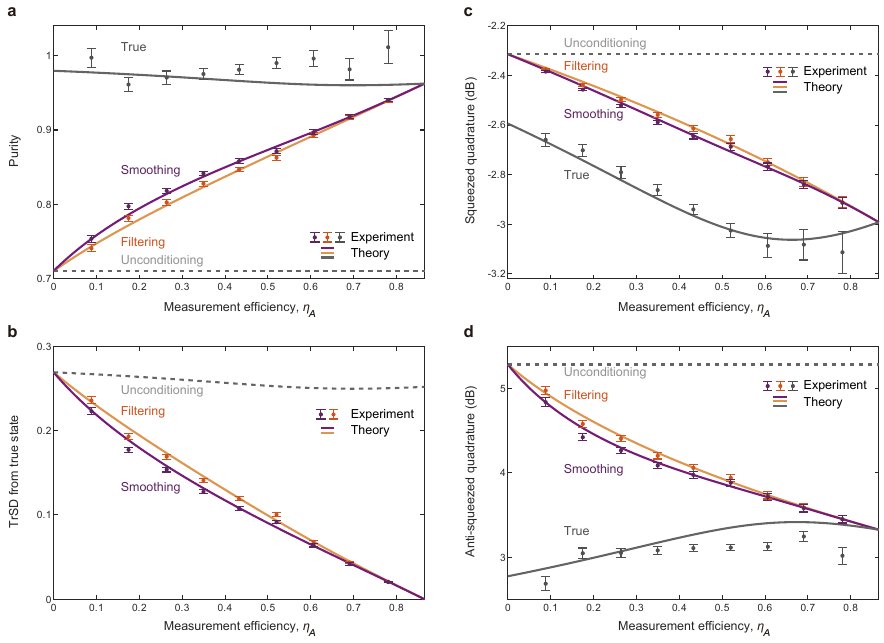}
\caption{\textbf{Measured quantities for state estimation as functions of measurement efficiencies.}
	Four state quantities are plotted as functions of Alice's measurement efficiency, $\eta_{\rm A}$, for the fixed values of $\theta_{\rm A} = \SI{65}{\degree}$ and $\theta_{\rm B} = \SI{135}{\degree}$. The quantities are:  \textbf{a}, state purity ${\cal P}(\rho_{\rm C})$; \textbf{b}, average TrSD from true states ${\cal D}(\rho_{\rm C})$; \textbf{c} and \textbf{d}, uncertainties in the squeezed and anti-squeezed quadratures in dB units to make the vacuum noise level \SI{0}{\decibel}, i.e., they are $10 \log {\cal S}(\rho_{\rm C})$ for squeezing and $10 \log {\cal A}(\rho_{\rm C})$ for anti-squeezing. The four quantities are shown for $\rho_{\rm C} =  \rho_{\rm T}, \rho_{\rm F}, \rho_{\rm S}, \rho_{\rm unc}$, which are true states, filtered states, smoothed states, and the unconditional state (equivalent to $\eta_{\rm A}=0$), respectively, where error bars correspond to one standard deviation.
 }
\label{fig:Result2}
\end{figure}

In order to analyze the improvement that smoothing offers over conventional filtering,
we define the following four quantities that can be applied to both filtered and smoothed estimated states: the state Purity, the average Trace-Squared Deviation (TrSD) from true states, the amount of quadrature squeezing and anti-squeezing. First is the state purity, which, for a Gaussian state, can be computed directly from the covariance by
${\cal P}(\rho_{\rm C}) \equiv 1/\sqrt{{\rm det}(2V_{\rm C}/\hbar)}$,
where ${\rm det}(\cdot)$ denotes a matrix determinant and the subscript C can be S or F for smoothing or filtering, respectively. The purity is the most straightforward to calculate: the ``smaller'' the covariance, the higher the purity.
We show, in \figref{Result2}a, for a fixed $\theta_{\rm A} = \SI{65}{\degree}$ and $\theta_{\rm B} = \SI{135}{\degree}$, the purity of the filtered ${\cal P}(\rho_{\rm F})$, smoothed ${\cal P}(\rho_{\rm S})$, and true states ${\cal P}(\rho_{\rm T})$ as a function of Alice's measurement efficiency. We can see that, in the low to intermediate range of $\eta_{\rm A}$, the smoothed state is purer than the filtered state, indicating that the information in the future record has allowed us to, retroactively, describe the dynamics of the OPO with less decoherence.
At high efficiencies, smoothing performs just as well as filtering. This is to be expected as when Alice's measurement efficiency increases, Bob's efficiency decreases and his record contains less information about the true state. Consequently, Alice's filtered state using her past-only record will converge to the true state as $\eta_{\rm A} \to \eta_{\rm tot}$ and $\eta_{\rm B} \to 0$. At $\eta_{\rm B} = 0$, the purity improvement gained by having the future record naturally vanishes.
Moreover, as we mentioned, the total efficiency was quite high in the experiment and we can now see that the true state purity ${\cal P}(\rho_{\rm T})$ is close to one.

Although it is clear that quantum state smoothing gives purer quantum states than filtering, this does not prove that the smoothed state is a better estimate of the underlying true state than the filtered state. To see that it is, we consider the second quantity: the average TrSD, \textit{i.e.}, ${\cal D}(\rho_{\rm C}) \equiv \mathbb{E}\!\left[ {\rm Tr}\{(\rho_{\rm C} - \rho_{\rm T})^2\}\right]$, which is a trace of a squared difference between an estimated quantum state $\rho_{\rm C}$ and a true state $\rho_{\rm T}$, averaged over all ensembles of true states as well as Alice's observed records (see Methods for the calculation of ${\cal D}$ from experimental data). The TrSD measure is chosen as it is the closest analogue of the mean-square-error cost function in classical estimation, and minimizing the expected TrSD cost function leads to the standard filtering and smoothing quantum state estimates\cite{CHANTASRI20211}. Interestingly, it has been shown theoretically\cite{laverick2021quantum} that the average TrSD for the smoothed (filtered) state reduces to the average difference between the true state's purity and the smoothed (filtered) state's purity. In our case, since the covariance of Gaussian states is deterministic, one need not worry about the latter average. That is, we can compute the average TrSD directly from ${\cal D}(\rho_{\rm S (F)}) = {\cal P}(\rho_{\rm T}) - {\cal P}(\rho_{\rm S(F)})$ (see the Supplementary Information for the full calculation). 
In \figref{Result2}b, we see that the average TrSD, ${\cal D}(\rho_{\rm S})$, of the smoothed state, is indeed smaller than that of the filtered state ${\cal D}(\rho_{\rm F})$. Thus, on average, the smoothed state is closer to the true state than the filtered state. As before, when the measurement efficiency increases, ${\cal D}(\rho_{\rm S})$ and ${\cal D}(\rho_{\rm F})$ converge and both eventually vanish at the maximum efficiency.
We also see good agreement between the experimental data and the theoretical value of the average TrSD, computed from the purity difference mentioned prior (theory lines).

While having a greater purity and smaller deviation from the true states is the hallmark of quantum state smoothing\cite{laverick2021quantum}, these are not the only quantities one may be interested in.
In certain quantum information processing tasks, the performance quality is determined solely by the amount of squeezing present\cite{walshe2019robust,asavanant2021time}.
We thus consider a third quantity: the amount of conditional state squeezing, i.e., the uncertainty in the most squeezed quadrature of the quantum state $\rho_{\rm C}$, denoted by ${\cal S}(\rho_{\rm C})$, which is quantified by the smaller eigenvalue of $(2/\hbar)V_{\rm C}$. In \figref{Result2}c, we show that smoothing provides states with less uncertainty in the squeezed quadrature (smaller ${\cal S}$) than filtering at low to intermediate values of $\eta_{\rm A}$. At high efficiencies, we see that the conditional squeezing improvement diminishes, for the same reason as the previous two measures. We also look at a fourth quantity, the amount of conditional anti-squeezing, i.e., the uncertainty in the most anti-squeezed quadrature, ${\cal A}(\rho_{\rm C})$, quantified by the larger eigenvalue of $(2/\hbar)V_{\rm C}$. In this case, we observe that the uncertainty in the anti-squeezed quadrature from smoothing is also less (smaller ${\cal A}$) than that of the filtered state. 
This is to be expected since the smoothed state is purer than the filtered state as well as having greater conditional squeezing.
Notably, unlike conventional purification (distillation of squeezing), which is achieved at the expense of squeezing (purity)\cite{glockl2006squeezed,dirmeier2020distillation}, our smoothing technique recovers squeezing and purity simultaneously.

\begin{figure}[t!]\centering
\includegraphics[scale= 1]{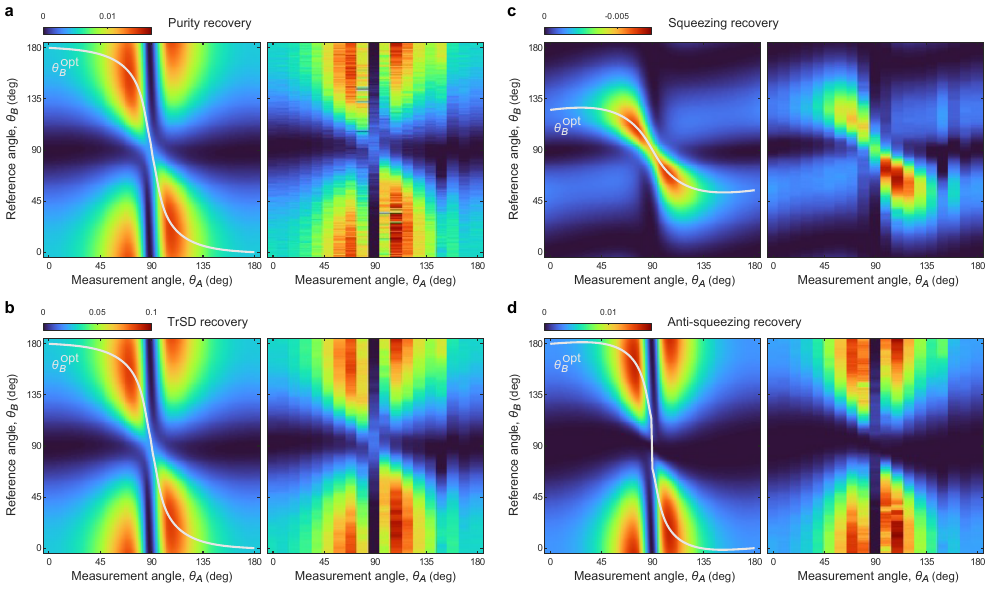}
\caption{\textbf{Improvement of smoothing over filtering for varying Alice's and Bob's measurement setups.}
\textbf{a}-\textbf{d}, The recovery quantities: the purity recovery ${\cal R}_{\cal P} \equiv {\cal P}(\rho_{\rm S})-{\cal P}(\rho_{\rm F})$, the TrSD recovery ${\cal R}_{\cal D} \equiv {\cal D}(\rho_{\rm F}) - {\cal D}(\rho_{\rm S})$, the squeezing recovery ${\cal R}_{\cal S} \equiv {\cal S}(\rho_{\rm F}) - {\cal S}(\rho_{\rm S})$, and the anti-squeezing recovery ${\cal R}_{\cal A} \equiv {\cal A}(\rho_{\rm F}) - {\cal A}(\rho_{\rm S})$, for varying $\theta_{\rm A}$ and $\theta_{\rm B}$. The left and right panels show the theoretical predictions and the corresponding experimental results, respectively. The grey curves indicate the optimal measurement angles for Bob $\theta^{\rm Opt}_{B}$ that maximize the recoveries for different Alice's measurement choice of $\theta_{\rm A}$. Alice's measurement efficiency for all data shown here is fixed at $\eta_{\rm A} = 0.43$.}
\label{fig:Result3}
\end{figure}

So far, we have only seen that smoothing performs better than filtering for \textit{fixed} measurement choices at $\theta_{\rm A} = \SI{65}{\degree}$ and $\theta_{\rm B} = \SI{135}{\degree}$ with different measurement efficiencies. Now we consider the entire space of possible measurement settings for both Alice and Bob, \textit{i.e.}, $\theta_{\rm A,\rm B} \in [0, \pi]$, for a fixed measurement efficiency $\eta_{\rm A} = 0.43$. Let us define the \textit{smoothing recovery} for the four quantities we have discussed in a way that yields positive values as measures of the improvement that smoothing offers over filtering. We thus have: the purity recovery ${\cal R}_{\cal P} \equiv {\cal P}(\rho_{\rm S})-{\cal P}(\rho_{\rm F})$, the TrSD recovery ${\cal R}_{\cal D} \equiv {\cal D}(\rho_{\rm F}) - {\cal D}(\rho_{\rm S})$, the squeezing recovery ${\cal R}_{\cal S} \equiv {\cal S}(\rho_{\rm F}) - {\cal S}(\rho_{\rm S})$, and the anti-squeezing recovery ${\cal R}_{\cal A} \equiv {\cal A}(\rho_{\rm F}) - {\cal A}(\rho_{\rm S})$, noting that purity is the only quantity that increases with improvement. In \figref{Result3}, we show the theoretical calculation and experimental data for the four recoveries. 

The theoretical calculation in \figref{Result3}a, b, c, and d (left panels) predicts that smoothing should provide improvements for all measurement choices on all recovery measures, with ${\cal R}_{\cal P, D, S, A} > 0$, even if only minor in some cases. 
Note that the theoretical purity recovery is exactly the same as the TrSD recovery, which is easy to verify from the relationship between them, and very similar to the anti-squeezing recovery.
Importantly, we see that each recovery metric's improvements depend on \textit{both} Alice's and Bob's measurement choice. The latter dependence can be solely attributed to the smoothed quantum state, as the filtered state can be computed without reference to Bob or his measurement choice. 
We have also included (grey) curves which show the \textit{optimal measurement settings} for Bob that maximize the respective recoveries for any given measurement setting for Alice. Interestingly, if we compare the optimal measurement settings found for the purity to those found for squeezing, there is only a single point that they intersect, $\theta_{\rm A} = \theta_{\rm B} = \SI{90}{\degree}$, where all improvements except ${\cal R}_{\cal S}$  are small. 

The experimental data for the recovery metrics in \figref{Result3}a, b, c, and d (right panels) are in good agreement with the theory. 
In the case of purity, we find a maximum recovery of $0.016\pm0.003$. When we compare this number to the total purity that we could have recovered, ${\cal P}(\rho_{\rm T}) - {\cal P}(\rho_{\rm F})$, we find that smoothing has recovered $10.3\%\pm 1.6\%$. Similarly, the maximum conditional squeezing recovery that smoothing yields is $0.006\pm0.002$, which is $7.6\%\pm 2.6\%$ of the total possible conditional squeezing, i.e., ${\cal S}(\rho_{\rm F}) - {\cal S}(\rho_{\rm T})$, that could be recovered.

\begin{figure}[t!]\centering
\includegraphics[scale=1]{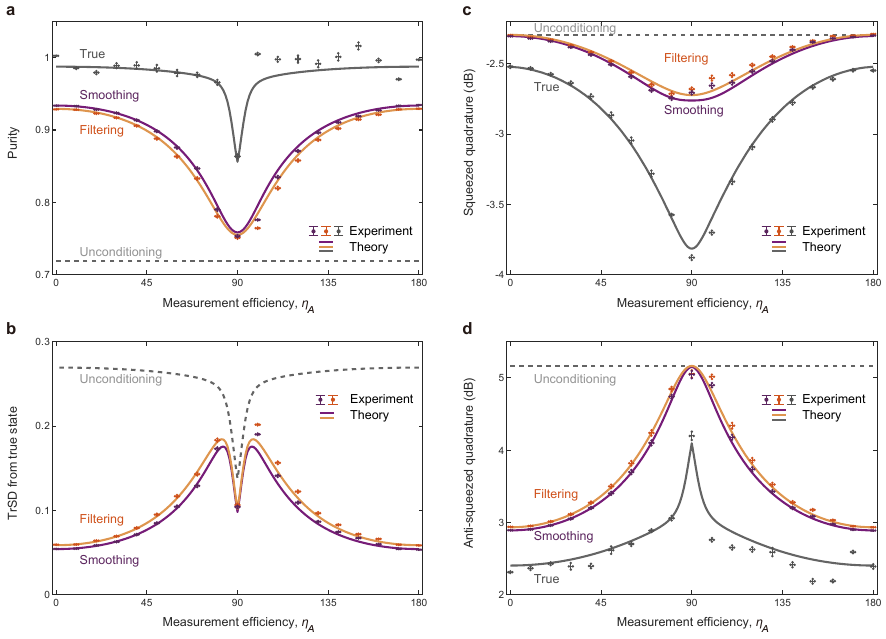}
\caption{\textbf{Measured quantities along the optimal measurement setting.}
The four quantities as in \figref{Result2} showing: \textbf{a}. state purity; \textbf{b}, average TrSD from true states; \textbf{c}, uncertainty in the squeezed quadrature (in dB units); and \textbf{d}, uncertainty in the anti-squeezed quadrature (in dB units). The quantities are evaluated along the grey curves in \figref{Result3}, which indicate the optimal measurement angle of Bob $\theta^{\rm Opt}_{B}$ for different values of Alice's measurement angles $\theta_{\rm A}$. Alice's measurement efficiency is fixed to $\eta_{\rm A}=0.43$.
}
\label{fig:Result4}
\end{figure}

Using the optimal measurement settings for Bob (grey curves) determined in the left panels of \figref{Result3}, we evaluate the four quantities of interest as functions of Alice's measurement setting $\theta_{\rm A}$, and plot them in \figref{Result4}. 
In \figref{Result4}a, we find that the purity is maximized at $\theta_{\rm A}=\SI{0}{\degree}$ (or \SI{180}{\degree}) and minimized at $\theta_{\rm A}=\SI{90}{\degree}$. We can see the similar curves for the average TrSD in \figref{Result4}b, which are simply upside-down versions of the purity curves, but with a sudden dip at $\theta_{\rm A}=\SI{90}{\degree}$. This arises from the dip in the purity of the true state  ${\cal P}(\rho_{\rm T})$ (which is discussed in the following paragraph), because of the relation ${\cal D}(\rho_{\rm S (F)}) = {\cal P}(\rho_{\rm T}) - {\cal P}(\rho_{\rm S(F)})$. The uncertainty in the anti-squeezed quadrature follows a similar, albeit inverted, trend to the purity, as shown in \figref{Result4}d. On the other hand, in \figref{Result4}c, the conditional squeezing is best at $\theta_{\rm A}=\SI{90}{\degree}$ and worst at $\theta_{\rm A}=\SI{0}{\degree}$ (or \SI{180}{\degree}).

\subsection*{Discussion}
\addcontentsline{toc}{subsection}{Discussion}
We now explain, qualitatively, the dependence of the results on the homodyne angles $\theta_{\rm A}$ and $\theta_{\rm B}$. We start with the trends in Fig.~\ref{fig:Result4} in the properties of the filtered state (which obviously only depends on $\theta_{\rm A}$). The filtered state has the highest purity when it has the least squeezing (greatest uncertainty in the squeezed quadrature). This is when  Alice measures the anti-squeezed quadrature $x$ ($\theta_{\rm A}=0^\circ$ or $180^\circ$), which has its smallest uncertainty at those points. Conversely, Alice's filtered state has the lowest purity when it is most squeezed. This is when she measures the squeezed quadrature $p$ ($\theta_{\rm A}=90^\circ$), as obtaining information about this quadrature also reduces the uncertainty in it, making it more squeezed than in steady-state. The correlation with purity is because squeezed states are more sensitive to loss, with the principle loss arising from Bob's channel, while less highly squeezed states are relatively more robust to loss. The smoothed state properties follow these trends, as even the best smoothing only partly compensates for the lost information in Bob's channel. Finally, the TrSD curves, for both smoothing and filtering, mostly follow the purity curves, as already discussed, with the noticeable deviation around $\theta_{\rm A}=90^\circ$ because of the dip in the purity of the true state in plot (a). This arises because, at this point, Bob is also measuring the $x$ quadrature, as per the grey line in Fig.~\ref{fig:Result3}a, leading to a maximally squeezed true state which is very sensitive to the residual (neither Alice nor Bob) loss of about $14\%$.

Turning now to the smoothing recoveries in Fig.~\ref{fig:Result3}, the patterns are quite complicated. The purity recovery ${\cal R}_{\cal P}$ for a very similar system was calculated\cite{laverick2021linear}, and closely matches our results here (a). In that paper, we sought an explanation for the shape of the optimal-measurement-choice curve (grey) in terms of measurement back-action. Here we note that its simplest feature, that for almost all choices of $\theta_{\rm A}$, Bob should measure close to the $x$ quadrature ($\theta_{\rm A}\approx 0^\circ$ or $180^\circ$), can be explained in another way. As noted above, by measuring the $x$ quadrature, Bob tends to reduce the uncertainty in $x$ (the anti-squeezed quadrature) only, creating a pure state that is less squeezed than for any other choice of his. That is, the pure state will be most robust to loss which, from Alice's perspective, is present because of Bob's channel. Thus she is best able to recover purity from that loss via smoothing when starting with a relatively loss-resistant pure state. This argument also explains why plot (d) of the anti-squeezing recovery ${\cal R}_{\cal A}$ looks almost identical to the purity recovery plot: the best way to increase the purity of the smoothed state is to reduce the uncertainty in the anti-squeezed quadrature. The ${\cal R}_{\cal P}$ plot also matched plot (d) of the TrSD recovery, ${\cal R}_{\cal D}$, as expected. 

By contrast with the above, plot (c) of the squeezing recovery, ${\cal R}_{\cal S}$, looks rather different. The most notable feature is that, in the region where significant squeezing recovery is possible, Bob's optimal measurement choice (grey curve) is not of the $p$ quadrature (the squeezed quadrature) as one might have guessed, but rather of a quadrature anti-correlated with Alice's (i.e., $\theta_{\rm B}\approx 180^\circ - \theta_{\rm A}$). This can be understood by considering the squeezing of the true state, as this choice by Bob leads to the greatest squeezing, as we predicted theoretically and verified experimentally (See Supplemental Information for the true state squeezing). Thus, once again, the smoothing power can be best understood from the properties of the true state, underlining the centrality of the true state to the method of quantum state smoothing.

In summary, we experimentally demonstrate that quantum state smoothing yields an estimated state with greater purity and higher squeezing than conventional filtering. Moreover, we show that the smoothed state provides a better estimate of the underlying true state than the filtered state. The restored purity, squeezing and the quality of the estimation depend on the measurement efficiency, measurement angle and estimation protocol, with smoothing out-performing filtering for a wide variety of measurement settings and at worst performs the same as the filtering. We find the dependence on the measurement angles, that is, the purity and the conditional squeezing of the estimated state cannot be maximized simultaneously.
Our techniques may be utilized for a variety of quantum information processing where the post-processing is allowed.
Although the detailed discussions of possible applications are beyond the scope of this work and will be investigated in a future study, one of the promising applications may be measurement-based quantum computation\cite{asavanant2021time,larsen2021deterministic} 
where measurements and the following feedforward operations enable quantum operations.
As the measurement records are stored during the processing, quantum state smoothing techniques may be embedded in the processing to restore the resources (\textit{i.e.}, entangled states) used in quantum computations and improve the performance.


\section*{Methods}
\addcontentsline{toc}{section}{Methods}
\subsection*{Dynamical equations for our OPO system}
\addcontentsline{toc}{subsection}{Dynamical equations for our OPO system}
Our experimental system is represented by quantum Langevin equations of Heisenberg picture operators describing the dynamics of the system's operators $\hat{\bf x} = \left( \hat x \,\,\, \hat p \right)\tp$, Alice's homodyne current $\hat{y}_{\rm A}$, and Bob's current $\hat{y}_{\rm B}$. The dynamical equations are\cite{laverick2019quantum}: 
\begin{subequations}
\begin{align}
	&d\hat{\mathbf{x}} = A\hat{\mathbf{x}}dt + B d\hat{\mathbf{v}}, \\
	&\hat{y}_{\rm A} dt = C_{\rm A} \hat{\bf x}dt + D_{\rm A} d\hat{\mathbf{v}},\\
	\label{eq:yA}
	&\hat{y}_{\rm B} dt = C_{\rm B} \hat{\bf x}dt + D_{\rm B} d\hat{\mathbf{v}},
\end{align}
\end{subequations}
where $d\hat{\bf v}$ represents the noise term, which is a vector of independent Wiener increments. The constant matrices are given by
\begin{subequations}
\begin{align}
	A &= \begin{pmatrix}
	-\gamma (1-\xi) & 0\\
	0 & -\gamma (1+\xi)
\end{pmatrix},\\
	B &= \sqrt{\frac{\hbar}{2}}\begin{pmatrix}
    \sqrt{\gamma_{L_c}}I_2 & \sqrt{\gamma_{T_c}}I_2 & 0_2 & 0_2 & 0_2\\
\end{pmatrix},\\
	C_{m} &= 2\sqrt{\frac{\gamma \eta_m}{\hbar}} M_m,\quad M_m = \begin{pmatrix}\cos\theta_m & \sin\theta_m\end{pmatrix}, \\
	D_m &= 
	\begin{pmatrix}
		0_2 M_m & d_2^{(m)}M_m & d_3^{(m)}M_m & d_4^{(m)}M_m & d_5^{(m)}M_m \\
	\end{pmatrix},
\end{align}
\end{subequations}
with $m = A,B$, the $2 \times 2$ identity matrix $I_2$, and the $2\times 2$ zero matrix $0_2$. The variables $\gamma_{T_c}$ and $\gamma_{L_c}$ denote, respectively, the cavity loss rate due to the output coupler and the intracavity loss rate, where $\gamma_{T_c} + \gamma_{L_c} = 2\gamma$. The particular forms of the $d^{(m)}_j$ constants and full derivations are given in the Supplementary Information. In this picture, we can see why classical smoothing fails for quantum systems. While the operator describing the homodyne measurements commutes with the future system operators $\hat{\bf x}$, the reverse is not true. That is, the future measurement operators do not necessarily commute with $\hat{\bf x}$ at past times. What this means is that one can predict, and subsequently filter, the system's behavior in the future, but one cannot retrodict (nor retrofilter) the system's behavior in the past.

\subsection*{Experimental parameters}
\addcontentsline{toc}{subsection}{Experimental parameters}
We set the normalized pump amplitude to $\xi = 0.70$ with \SI{600}{\milli\watt} pumping.
The half-width at half maximum of our cavity is $\gamma/2\pi = \SI{5.0}{\mega\hertz}$.
The total measurement efficiencies for the homodyne detections are $\eta_{\rm A} = 0.865T$ and $\eta_{\rm B} = 0.863(1-T)$, where $T$ is the energy transmittance of the variable BS.
The measurement angles $\theta_{\rm A}$ and $\theta_{\rm B}$ are for Alice's and Bob's homodyne detections.
Note that $\hbar$ is normalized to 1 in the experimental results. 
See Supplementary Information for details of experiments.

\subsection*{Quantum state smoothing for linear Gaussian system}
\addcontentsline{toc}{subsection}{Quantum state smoothing for Linear Gaussian system}
For a linear Gaussian system, a quantum state is estimated in a similar way to the classical Kalman filtering and smoothing\cite{laverick2019quantum}.
A covariance of a quantum state is related to an estimation error matrix.
In our study, the covariance of the true state, $V_{\rm T}$, the filtered covariance, $V_{\rm F}$, and the smoothed covariance, $V_{\rm S}$, can be solved deterministically using the following equations:
\begin{subequations}
\begin{align}
	\frac{d V_{\rm T}}{dt} =&\, AV_{\rm T} +V_{\rm T} A\tp + \hbar \gamma I_2 -K_{\rm A}[V_{\rm T}]K_{\rm A}[V_{\rm T}]\tp
	-K_{\rm B}[V_{\rm T}]K_{\rm B}[V_{\rm T}]\tp,\\
	\frac{d V_{\rm F}}{dt} =&\,AV_{\rm F} +V_{\rm F} A\tp + \hbar \gamma I_2
	-K_{\rm A}[V_{\rm F}]K_{\rm A}[V_{\rm F}]\tp, \\
	-\frac{d \halo{\Lambda}_{\rm R}}{dt} =&\,\halo{\Lambda}_{\rm R}\bar{A} +\bar{A}\tp \halo{\Lambda}_{\rm R} - \halo{\Lambda}_{\rm R}\bar{Q} \halo{\Lambda}_{\rm R} + C_{\rm A}\tp C_{\rm A}^{},
\end{align}
\end{subequations}
with a relationship among all the covariance
\begin{align}
	V_{\rm S} = V_{\rm T} + (V_{\rm F}- V_{\rm T})\left[I_2 + \halo{\Lambda}_{\rm R} (V_{\rm F} - V_{\rm T})\right]^{-1},
\end{align}
where $\bar{A} = A - K_{\rm A}[V_{\rm T}] C_{\rm A}$ and $\bar{Q} = K_{\rm B}[V_{\rm T}]K_{\rm B}[V_{\rm T}]\tp$, using the optimal Kalman gain given by $K_{\rm A,B}[V] = \left(V -\frac{\hbar}{2}I_2\right) C_{\rm A,B}\tp$. 
We note that $V_{\rm T}$ is mathematically equivalent to a covariance of a `filtered' state that is conditioned on both Alice's and Bob's measurements with no reference to the remaining experimental loss. The matrix $\halo{\Lambda}_{\rm R}$ is related to the retrofiltered covariance of a positive-operator-valued-measure (POVM) element describing the future measurement record, $V_{\rm R}$, via the relation\cite{laverick2019quantum}: $\halo{\Lambda}_{\rm R} = (V_{\rm R} + V_{\rm T})^{-1}$. 
The mean of the true state, $\ex{\hat{\bf x}}_{\rm T}$, is calculated using Alice's and Bob's measurement records, $\mathbf{y}_{\rm A}$ and ${\bf y}_{\rm B}$ (note the lack of hat accents for classical variables), from
\begin{align}
	\frac{d\ex{\hat{\bf x}}_{\rm T}}{dt} &= A\ex{\hat{\bf x}}_{\rm T}  + K_{\rm A} [V_{\rm T}] \left( \mathbf{y}_{\rm A} - C_{\rm A} \ex{\hat{\bf x}}_{\rm T} \right) 
	+ K_{\rm B} [V_{\rm T}] \left( \mathbf{y}_{\rm B} - C_{\rm B} \ex{\hat{\bf x}}_{\rm T} \right).
\end{align}
On the other hand, the mean of the filtered state, $\xf$, and that of the smoothed state, $\xs$, are estimated conditioned on the acquired Alice's measurement records, $\mathbf{y}_{\rm A}$, as
\begin{subequations}
\begin{align}
	\frac{d\ex{\hat{\bf x}}_{\rm F}}{dt} &= A\ex{\hat{\bf x}}_{\rm F}  + K_{\rm A} [V_{\rm F}] \left( \mathbf{y}_{\rm A} - C_{\rm A} \ex{\hat{\bf x}}_{\rm F} \right),  \\
	-\frac{d \halo{\mathbf{z}}_{\rm R}}{dt} &= \left(\bar{A}\tp - \halo{\Lambda}_{\rm R} \bar{Q}\right) \halo{\mathbf{z}}_{\rm R} 
	+ \left(C_{\rm A}\tp - \halo{\Lambda}_{\rm R} K_{\rm A}[V_{\rm T}]\right) \mathbf{y}_{\rm A}, \\
	\ex{\hat{\bf x}}_{\rm S} &= \left[I_2 + (V_{\rm F} - V_{\rm T}) \halo{\Lambda}_{\rm R} \right]^{-1} \left[ \ex{\hat{\bf x}}_{\rm F} + (V_{\rm F} - V_{\rm T}) \halo{\mathbf{z}}_{\rm R}\right],
\end{align}
\end{subequations}
where $\halo{\mathbf{z}}_{\rm R}  = \halo{\Lambda}_{\rm R}\ex{\hat{\bf x}}_{\rm R}$. See the Supplementary Information for why it is necessary to use $\halo{\mathbf z}_{\rm R}$ as opposed to $\ex{\hat{\bf x}}_{\rm R}$.
The filtered state is a function of Alice's measurement angle $\theta_{\rm A}$, while the smoothed state is a function of both Alice's angle $\theta_{\rm A}$ and Bob's measurement angle $\theta_{\rm B}$.

\subsection*{Data acquisition}
\addcontentsline{toc}{subsection}{Data acquisition}
We repeated \SI{50}{\micro\second} data acquisition 600~times for each $T$ for the data shown in \figref{Result1}--\figref{Result2}, and \SI{10}{\milli\second} data acquisition 30~times for each $\theta_{\rm A}$ for the data shown in \figref{Result3} and \figref{Result4}.
We locked the phases $\theta_{\rm A}$ and $\theta_{\rm B}$ in \figref{Result1}--\figref{Result2}.
On the other hand, in \figref{Result3} and \figref{Result4}, we scanned the phase $\theta_{\rm B}$ around \SI{140}{\hertz} and recorded the error signals of $\theta_{\rm B}$.
We determined the phase $\theta_{\rm B}$ from the error signals, divided the dataset into \SI{1}{\degree} blocks, and calculated the filtered and smoothed estimates for each block.

\subsection*{Reconstruction of covariance}
\addcontentsline{toc}{subsection}{Reconstruction of covariance}
In order to verify our experiment, we reconstructed the covariances using the mean square error matrices between the estimates with and without Bob's measurement records.
The filtered and smoothed covariances conditioned on Alice's measurement records are reconstructed as follows,
\begin{subequations}
\begin{align}
	 \tilde V_{\rm F} &= 
	 \mathbb{E}\left[ (\ex{\hat{\bf x}}_{\rm F} -  \ex{\hat{\bf x}}_{\rm T}) (\ex{\hat{\bf x}}_{\rm F} -  \ex{\hat{\bf x}}_{\rm T} ) \tp\right]   + V_{\rm T}, \\
	\tilde V_{\rm S} &=
	\mathbb{E}\left[ (\ex{\hat{\bf x}}_{\rm S} -  \ex{\hat{\bf x}}_{\rm T}) (\ex{\hat{\bf x}}_{\rm S} -  \ex{\hat{\bf x}}_{\rm T} ) \tp\right]   + V_{\rm T},
\end{align}
\end{subequations}
where ${\mathbb E}$ denotes an ensemble average over the measurement records. In these expressions we use $V_{\rm T}$ as calculated from the theory. This ensures that we can compare quantities related to the filtered and smoothed states without having to worry about the common noise resulting from errors in the experimentally derived $\tilde{V}_{\rm T}$. The latter, which we used when we plotted properties of the true state, is defined by 
\begin{align}
    \tilde{V}_{\rm T} &= V_{\rm unc} - {\rm Cov}(\ex{\hat{\bf x}}_{\rm T},\mathbf{0}).
\end{align}
where $V_{\rm unc}$ is the covariance of the unconditional state and $\mathbf{0}$ represents a vector of zeros.
These reconstructed covariances are then used to calculate the purities, conditional squeezing and conditional anti-squeezing shown in \figref{Result2}--\figref{Result4}.

\subsection*{Average Trace-Squared Deviation (TrSD)}
\addcontentsline{toc}{subsection}{Average Trace-Squared Deviation (TrSD)}
Although average TrSD is theoretically equivalent to the purity difference, as shown in the Supplementary Information, we compute it without reconstructing the covariances, to independently verify the experimental consistency.
We evaluate the average overlap between Gaussian distributions defined by the estimates with and without Bob’s measurement records:
\begin{align}
    {\cal D}(\rho_{\rm C}) = \frac{\hbar}{2}\left(\sqrt{|V_{\rm T}|^{-1}}+\sqrt{|V_{\rm C}|^{-1}}\right) - 4\pi \hbar \, \mathbb{E}\left[ g(\ex{\hat{\bf x}}_{\rm T} - \ex{\hat{\bf x}}_{\rm C}, V_{\rm T} + V_{\rm C}) \right],
\end{align}
where $g({\bf x}, V) = (2\pi\sqrt{|V|})^{-1} \exp\left(-\frac{1}{2}{\bf x}^\top V^{-1} {\bf x} \right)$ is a Gaussian function. The subscript ${\rm C}$ indicates filtering (F) or smoothing (S), and ${\rm T}$ denotes the true state.

%

\section*{Acknowledgments}
\addcontentsline{toc}{section}{Acknowledgements}
This work was supported financially by the Australian Research Council Centres of Excellence Scheme No.~CE170100012.
S.Y. acknowledges the funding from Japan Society for the Promotion of Science (JSPS) KAKENHI (Grant No. 24K06930), and D.D. acknowledges the support from the Australian Research Council Future
Fellowship Funding Scheme under Project FT220100656.

\section*{Author contributions}
\addcontentsline{toc}{section}{Author contributions}
K.T.L., A.C., and H.M.W. conceived the original idea.
S.Y., D.M., and H.Y. planned, designed, built, conducted the experiment, and acquired the data.
S.Y. formulated the model for the experiment with assistance from D.M., Q.Y., D.D., W.A., and H.Y.
S.Y. prepared the code and analyzed the data with support from K.T.L., A.C., H.M.W., and H.Y.
All authors contributed to writing the manuscript.

\section*{Competing interests}
\addcontentsline{toc}{section}{Competing interests}
The authors declare no competing interests.

\section*{Additional information}
\addcontentsline{toc}{section}{Additional information}
Supplementary Information is available for this paper.



\bibliography{QSS_Bib}

\clearpage

\renewcommand{\thefigure}{S\arabic{figure}}
\setcounter{figure}{0}
\renewcommand{\thefigure}{S\arabic{figure}}
\setcounter{figure}{0}

\begin{spacing}{2}
	\begin{center}
{\LARGE Supplementary Information for Tracking Quantum Dynamics in an Optical Cavity for Recovering Purity and Squeezing via Quantum State Smoothing} 
	\end{center}
\end{spacing}

\section{Experimental details}
\begin{figure}[!h]\centering
\includegraphics[width=1\textwidth]{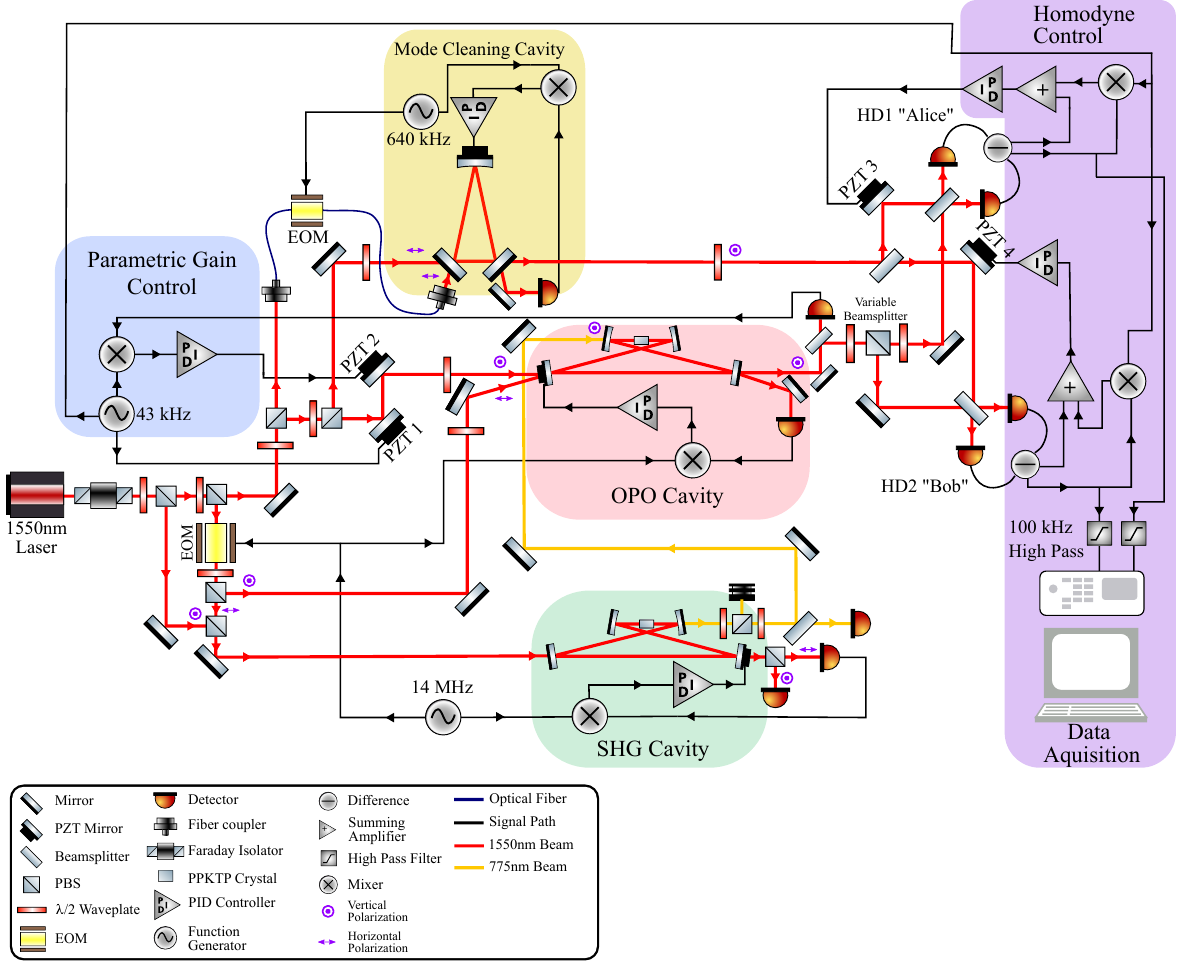}
\caption{\label{fig:Setup}A schematic of the full experimental setup, including the optical system and the control paths used to operate the experiment and take measurements. The legend in the bottom left shows what each symbol represents in the setup.}
\end{figure}

The schematic of our optical system and control paths is shown in \figref{Setup}.
The following sections describe the optics and control systems of the experiment in detail. 

\subsection{\label{sup:optics}Optical system}
A continuous-wave \SI{1550}{\nano\meter} laser is used as the optical source for this experiment. The laser power is around \SI{2.7}{\watt} at the source, which is split into the numerous paths as illustrated in \figref{Setup}.

\figureref{Setup} shows the different optical sub-systems used in the experiment. An optical parametric oscillator (OPO) is used to prepare the quantum state (shown in red). A mode cleaning cavity (MCC) is employed to prepare a local oscillator (LO) beam that is used for homodyne detection (shown in yellow). The OPO is pumped by a \SI{775}{\nano\meter} beam that is produced via second harmonic generation (SHG) from a cavity (shown in green). The squeezed beam from the OPO is split using a variable beam splitter, which is a combination of a polarizing beam splitter (PBS) and half waveplates used to control the splitting ratio. Both the transmitted and reflected beams are interfered with LO beams and then measured with homodyne detectors. 

Each cavity requires a locking beam and a probe beam.
Each locking beam is modulated using an electro-optic modulator (EOM) to generate sidebands that are used to lock the cavities using the Pound-Drever-Hall (PDH) technique\cite{drever1983laser}. Each cavity features a mirror mounted to a PZT, which is used to actively tune the cavity length to match the laser wavelength using feedback from the PDH error signal. 

The MCC is configured as a three mirror ring cavity with the round-trip length of \SI{800}{\milli\meter} which allows for spatial separation of the fields propagating clockwise and anti-clockwise at the output. 
The cavity is crafted out of a solid block of aluminum to provide stability, and is sealed to minimize loss. For this experiment the LO power is set to \SI{1}{\milli\watt}.

The SHG cavity is a bow-tie cavity with a periodically poled potassium-titanyl-phosphate (PPKTP) crystal mounted inside. The interaction between the intracavity field and the PPKTP crystal results in second harmonic generation, which produces a beam at the double frequency of the input beam (\SI{775}{\nano\meter}). The crystal is mounted on a 3-axis translation stage which allows for fine control of its position. The crystal has the size of 11(L)$\times$2.5(W)$\times$1(H)\SI{}{\milli\meter\cubed} and two wedged surfaces (the surfaces that the beam is transmitted through) to make the shape of a trapezium. This shape allows for the effective length of the crystal that interacts with the beam to be fine tuned by moving the translation stage perpendicular to the beam. The wedged surfaces also minimize the interference of fields that are reflected off the crystal surface.

The conversion efficiency of the PPKTP crystal is temperature dependent; hence, for a given input power, the amount of light generated at \SI{775}{\nano\meter} will change according to the crystal temperature. A temperature controller is used to heat the crystal inside the cavity so that we can tune this conversion efficiency. As can be seen in \figref{Setup}, the locking beam and probe beam for the SHG cavity use horizontal and vertical polarization, respectively (the beam polarization is indicated by a symbol in \figref{Setup} as shown in the legend at the bottom of the figure). These two beams are combined on a PBS for injection into the cavity through the same port. The beams are separated by a PBS at the output port of the SHG cavity, where two different detectors are used for cavity locking and for monitoring the probe power. Since the crystal is birefringent, the crystal position and temperature can both be used to tune the resonance frequencies of the two different polarized beams to match. The phase matching temperature of the crystal has a very broad maximum which allows us to fine tune the resonance with the temperature controller without sacrificing conversion efficiency. The output power of the SHG is controlled using a variable BS (the power of \SI{600}{\milli\watt} is used for the results shown in this paper), and is monitored by a detector which measures a pick-off beam in the pump path.

\begin{table}[b!]\centering
\caption{\label{tab:tableOPO} Parameters regarding OPO.}
\begin{tabularx}{\textwidth}{Xcc} \hline\hline
Cavity parameters & Variable & Value \\ \hline
Parametric gain & $G_+$ & 11.4 \\
Normalized pump amplitude & $\xi = 1-1/\sqrt{G_+}$ & 0.70 \\
Optical round-trip length & $l$ & \SI{489}{\milli\meter} \\
Output coupler & $T_c$ & 10.0\% \\
Intracavity loss & $L_c$ & 0.282\% \\
Half width at half maximum & $f_\mathrm{HWHM} = \gamma/(2\pi)=c (T_c+L_c)/(4\pi l)$ & \SI{5.0}{\mega\hertz} \\
Escape efficiency & $\eta_c = T_c /(T_c+L_c)$ & 97.3\% \\\hline\hline
\end{tabularx}
\caption{\label{tab:tableEff} Parameters regarding efficiency.}
\begin{tabularx}{\textwidth}{Xccc} \hline\hline
	Efficiency parameters			&	Variable	 & Alice & Bob \\ \hline
	Visibility 	& $\zeta_m$		&	99.2\%		&	99.3\%	\\
	Propagation loss	& $1-\eta_m^\mathrm{(prop)}$ & 8.4\%	& 8.8\% \\
	Shot noise clearance (\SI{1}{\mega\hertz} -- \SI{5}{\mega\hertz})	& $1-\eta_m^\mathrm{(elec)}$ &  $\SI{-25.4}{\decibel}$	& 	$\SI{-25.6}{\decibel}$ \\
	Quantum efficiency of photodiode & $\eta_m^\mathrm{(PD)}$ & 99\% & 99\%  \\
	Total efficiency without $T$ & 
	$\eta_{m0}^{}=\eta_c^{} \zeta_m^2 \eta_m^\mathrm{(prop)}\eta_m^\mathrm{(elec)}\eta_m^\mathrm{(PD)}$
	& 86.5\% & 86.3\% 
\\\hline\hline
\end{tabularx}
\end{table}

The OPO is another bow-tie cavity with a PPKTP crystal, which is used to generate squeezed states via spontaneous parametric down-conversion. The mechanical structure of the OPO is the same as the SHG cavity. The designed optical round-trip length of the cavity is \SI{489}{\milli\meter}. The measured transmission ratio of the output coupler is 10.0\%, and the intracavity loss of the OPO is 0.282\%. The pump beam used for this non-linear optical process is the \SI{775}{\nano\meter} beam created from the SHG cavity as described above.
As with the SHG cavity, the crystal inside the OPO is positioned on a translation stage and has wedged surfaces. The conversion efficiency of the crystal is temperature dependent, and a temperature controller is used to heat the crystal in the same manner as the SHG cavity. For the OPO, the locking beam with \SI{1}{\milli\watt} and probe beam use horizontal and vertical polarization, respectively, as indicated in \figref{Setup}. These beams are injected through the same mirror, but are injected at different angles so that they propagate in opposite directions around the cavity. This allows spatial separation of the locking and probe beams as they leave the cavity, meaning that the locking beam does not contaminate the OPO probe output, and can be detected separately. The probe power is \SI{80}{\nano\watt} at the output so that the interference signal with an LO beam does not saturate at the homodyne detector. As with the SHG cavity, the temperature and crystal position can both be tuned to match the resonance frequencies of the locking and probe beams. 
For the measurements presented in this paper, the pump beam power is set to \SI{600}{\milli\watt}, resulting in a parametric gain of $G_+ = 11.4$. 

A 99\% reflective beam splitter is used to pick off 1\% of the OPO probe beam output. This measured probe output is fed back to the relative phase of the probe beam through PZT 2 (see \figref{Setup}) in the OPO input path. A PBS is used along with a half waveplate to split the OPO output to the two homodyne detectors. The homodyne detectors are custom made with a quantum efficiency specification of 99\%. The detector consists of a transimpedance amplifier with \SI{39}{\kilo\ohm}, a cascade buffer amplifier, and a high-pass filter with the cutoff frequency of \SI{8}{\kilo\hertz} in between them. The clearance, the ratio between shot noise and detector electronic noise, is about $\SI{-25}{\decibel}$ on average in the range from \SI{1}{\mega\hertz} to \SI{5}{\mega\hertz} with \SI{1}{\milli\watt} LO power. The LO and OPO outputs are mode-matched to each other, with a typical visibility of 99.3\%. The propagation loss in these detection paths from the OPO output has been measured at $8.4\%$ for Alice and $8.8\%$ for Bob.

The experimental parameters used for quantum state smoothing are listed in Table~\ref{tab:tableOPO} and Table~\ref{tab:tableEff}. Here $c=3\times10^8$~\SI{}{\meter/\second} is the speed of light in the vacuum. Note that $\eta_{m0}$ corresponds to the total measurement efficiency without the transmission ratio of the variable beamsplitter $T$,
\begin{align}
\eta^{}_{A0} = \frac{\eta^{}_{\rm A}}{T}, \qquad
\eta^{}_{B0} = \frac{\eta^{}_{\rm B}}{1-T},
\end{align}
where $\eta^{}_{\rm A}$ and $\eta^{}_{\rm B}$ are defined by Eq.~\!\eqref{eq:measEff}.

\subsection{\label{sec:control}System control}
As mentioned in Section \ref{sup:optics}, the three cavities are locked using the PDH technique with modulated locking beams. These control loops can be seen in \figref{Setup}, with the control loops for the SHG cavity, OPO, and MCC highlighted in green, red, and yellow, respectively. 

A modulation frequency of \SI{14}{\mega\hertz} is used for the free space EOM in the bottom left of \figref{Setup} to produce sidebands for locking the SHG and OPO cavities. For the fiber EOM, a modulation frequency of \SI{640}{\kilo\hertz} is used for modulating the locking beam of the mode cleaning cavity. 

The parametric gain is controlled by modulating a PZT mirror with the frequency of \SI{43}{\kilo\hertz} in the input path of the OPO probe beam (PZT 1 in \figref{Setup}). This modulation signal is then mixed with a detected signal at the output of the OPO cavity to create an error signal that is injected into a servo which actuates on PZT 2 in the OPO probe input path. This actuation controls the relative phase of the probe and pump beams in the OPO which controls the parametric gain. This configuration is illustrated in \figref{Setup}, with control paths for the parametric gain control highlighted in blue.

For each of the homodyne detectors, a PZT mirror is used in the LO path to control the relative phase between the squeezed beam and the LO beam when they interfere (PZT 3 is used for `Alice' and PZT 4 is used for `Bob'). By controlling this phase it is possible to tune the measurement angle of the detector. The error signal for this control loop is created by summing together two components: the DC signal from the homodyne detector, and the demodulated AC signal. These components individually can be used as an error signal to lock to either a squeezing (\SI{90}{\degree} measurement angle) or anti-squeezing (\SI{0}{\degree} angle) measurement. By combining them it is possible to lock the homodyne detector to any angle by simply tuning the relative gains of these two signals. \figureref{Setup} shows this control configuration highlighted in purple. The measurement control loop has a phase noise of around $\pm \SI{2}{\degree}$. 

The signals from the homodyne detectors are measured using a Tektronix DPO7104 Digital Phosphor Oscilloscope. These signals are passed through the fifth-order Butterworth high-pass filter with the cutoff frequency of \SI{100}{\kilo\hertz} before measurement to remove a signal that is observed from the \SI{43}{\kilo\hertz} PZT frequency.  These measurements are recorded at \SI{1}{\giga S/\second} for a duration of \SI{10}{\milli\second} or \SI{50}{\micro\second} to be used as data for estimation of the state dynamics.

\section{Theoretical calculations}
\subsection{\label{sec:detailedmodel}Detailed experimental model}
\begin{figure}[!b]\centering
	\includegraphics[scale=1.5]{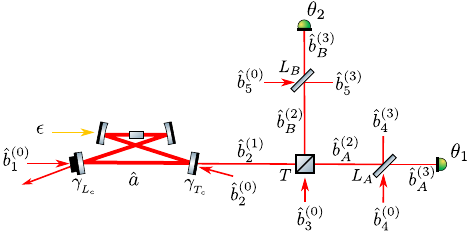}
	\caption{The schematics of our detailed experimental model. An OPO with injected pump amplitude $\epsilon$. Vacuum inputs to the system are shown as $\hat{b}^{(0)}_j (j=1,\dots,5)$.
    Homodyne measurements are shown by the measurement phases $\theta_1$ and $\theta_2$. $\gammaLc$ and $\gammaTc$ are the decay rate given by intracavity loss $L_c$ and output coupler transmission $T_c$. $L_{\rm A}$ and $L_{\rm B}$ represent measurement losses. $T$ is the transmission ratio of a variable beam splitter. 
    }
	\label{fig:detailedModel}
\end{figure}

In \figref{detailedModel}, we show our experimental model of the OPO system involving all the necessary parameters required. The annihilation (creation) operator describing the intra-cavity mode is denoted by $\hat{a}$ ($\hat{a}^\dagger$). The intra-cavity loss is modelled by a partially-reflective mirror (lower left in the cavity) with transmissivity $L_c$. There is an additional partially-reflective mirror, with transmissivity $T_c$, whose output is used for measurement. The decay rate of the cavity mode due to the intra-cavity loss and the output coupler, respectively, are given by 
\begin{equation}
\gammaLc = \frac{c}{l} L_c\qquad {\rm and}\qquad \gammaTc = \frac{c}{l} T_c\,, 
\end{equation}
where $c$ is the speed of light and $l$ is the round-trip length of the cavity.

The Hamiltonian of the degenerate OPO is given as\cite{collett1984squeezing},
\begin{align}
	\hat{\mathcal{H}} = \hbar \omega_c \hat a^\dagger \hat a + \frac{i\hbar}{2} \left[
	\epsilon \ee^{-i\omega_p t} (\hat a^\dagger)^2 - \epsilon^* \ee ^{i\omega_pt}\hat a^2\right],
\end{align}
where $\omega_c$ is the resonance frequency of the cavity, $\omega_p = 2\omega_c$ is the frequency of the beam pumping the cavity which has an amplitude of $\epsilon \in {\mathbb R}$. Moving to a rotating frame with respect to $\hbar\omega_c\hat a^\dagger \hat a$, the Hamiltonian becomes
\begin{equation}
\hat{\mathcal{H}}^{\rm RF} = \frac{i\hbar}{2} \epsilon[(\hat{a}^\dagger)^2 - \hat{a}^2]\,.
\end{equation}
Using standard input-output theory\cite{collett1984squeezing, gardiner2004quantum}, we obtain the Heisenberg equation of motion for the cavity mode,
\begin{equation}\label{Langevin}
\begin{split}
\frac{d\hat{a}}{dt} &= -\frac{i}{\hbar}[\hat{a}(t),\hat{\mathcal{H}}^{\rm RF}] - \gamma \hat{a}(t) + \sqrt{\gammaLc}\hat{b}_1^{(0)}(t) + \sqrt{\gammaTc}\hat{b}_2^{(0)}(t) \\
& = \epsilon \hat{a}^{\dagger}(t) - \gamma \hat{a}(t) + \sqrt{\gammaLc}\hat{b}_1^{(0)}(t) + \sqrt{\gammaTc}\hat{b}_2^{(0)}(t)\,,
\end{split}
\end{equation}
where
\begin{equation}
\gamma = \frac{1}{2}(\gammaLc + \gammaTc)\,.    
\end{equation}
The operators $\hat{b}_j^{(0)}$ are annihilation operators describing the input field at each partially-reflective mirror. The field at each mirror is assumed independent, with $[\hat{b}_j^{(0)}(t), \hat b^{(0)}_{j'}(t')] = 0$ and $[\hat{b}_j^{(0)}(t),(\hat b^{(0)}_{j'}(t'))^\dagger] = \delta(t - t')\delta_{j,j'}$ (see \figref{detailedModel} for our numbering conventions for $j$), and the superscript zero denotes a vacuum field. It is also convenient to introduce the infinitesimal noise operator $d\hat{B}_j^{(0)}(t) = \hat{b}_j^{(0)}(t) dt$, where $[d\hat{B}_j^{(0)}(t),(d \hat{B}_j^{(0)}(t'))\dg] = \delta(t - t')\dd t^2$. With this noise operator introduced, dynamics of the cavity is described by the stochastic differential equation
\begin{equation}
    d \hat{a} = \epsilon \hat{a}\dg dt - \gamma \hat{a} dt + \sqrt{\gammaTc}d\hat{B}_1^{(0)} + \sqrt{\gammaLc}d\hat{B}_3^{(0)}\,,
\end{equation}
where, for simplicity, we have dropped the time argument, and from this point on-wards unless necessary.
With this, we have the dynamics of the system. Before we write these dynamics in a standard form involving the canonical position and momentum operators, we will address Alice's and Bob's measurements.

To determine the operators that Alice and Bob are (effectively) measuring, we need to know how the output field $\hat{b}_{2}^{(1)}$ is related to the intra-cavity field. From the input-output calculation earlier, we have
\begin{equation}
\hat{b}^{(1)}_{2} = \sqrt{\gammaTc} \hat{a} - \hat b_{2}^{(0)}\,.
\end{equation}
From \figref{detailedModel}, this output field encounters a variable beam splitter, with transmissivity $T$, which divides the output into a portion that is observed by Alice and another by Bob. The output fields from this beam splitter are given by the transformation
\begin{equation}
	\begin{pmatrix}
		\hat b_{\rm A}^{(2)} \\
		\hat b_{\rm B}^{(2)} 
	\end{pmatrix}
	=
	\begin{pmatrix}
		\sqrt{T} & -\sqrt{1-T}\\
		\sqrt{1-T} & \sqrt{T} 
	\end{pmatrix}
	\begin{pmatrix}
		\hat b_2^{(1)} \\
		\hat b_3^{(0)} 
	\end{pmatrix}\,,
\end{equation}
where the second input to the beam splitter is an independent vacuum field with annihilation operator $\hat{b}_3^{(0)}$. The output fields are then sent to Alice and Bob who each perform a homodyne measurement. However, there is additional loss that is experienced along each path. This is modelled in the standard way by introducing a beam splitter in each arm whose reflectivity corresponds to the fraction of the field lost, $L_m$. Thus, the fields that Alice and Bob, respectively, obtain are
\begin{align}
	\begin{pmatrix}
		\hat b_{\rm A}^{(3)} \\
		\hat b_4^{(3)} 
	\end{pmatrix}
	&=
	\begin{pmatrix}
		\sqrt{1-L_{\rm A}} & -\sqrt{L_{\rm A}}\\
		\sqrt{L_{\rm A}} & \sqrt{1-L_{\rm A}} 
	\end{pmatrix}
	\begin{pmatrix}
		\hat b_{\rm A}^{(2)} \\
		\hat b_4^{(0)} 
	\end{pmatrix},\\
	\begin{pmatrix}
		\hat b_{\rm B}^{(3)} \\
		\hat b_5^{(3)} 
	\end{pmatrix}
	&=
	\begin{pmatrix}
		\sqrt{1-L_{\rm B}} & -\sqrt{L_{\rm B}}\\
		\sqrt{L_{\rm B}} & \sqrt{1-L_{\rm B}} 
	\end{pmatrix}
	\begin{pmatrix}
		\hat b_{\rm B}^{(2)} \\
		\hat b_5^{(0)} 
	\end{pmatrix}.
\end{align}

Note, $1-L_m$ corresponds to the measurement efficiency \textit{without} the escape efficiency from a cavity, and is determined via
\begin{align}
    1-L_m = \zeta_m^2 \eta^\mathrm{(prop)}_m \eta^\mathrm{(elec)}_m \eta^\mathrm{(PD)}_m,
\end{align} 
with the efficiency parameters: $\zeta$, visibility between a probe and an LO; $\eta^\mathrm{(prop)}$, optical propagation efficiency; $1-\eta^\mathrm{(elec)}$, the ratio of electrical noise power to shot noise power; and $\eta^\mathrm{(PD)}$, quantum efficiency of the photodiodes.

The overall measurement efficiency for both Alice and Bob, including the escape efficiency of the cavity, is given by
\begin{equation}
\label{eq:measEff}
\eta_{A} = \eta_c (1-L_{\rm A}) T\qquad{\rm and}\qquad \eta_{\rm B} = \eta_c (1-L_{\rm B}) (1-T)\,,    
\end{equation}
with $\eta_c = \gammaTc/(\gammaTc + \gammaLc)$.
After computing all the beam splitters along Alice's and Bob's paths, the fields that they receive in terms of the intra-cavity mode are 
\begin{align}
\hat{b}_{\rm A}^{(3)} &= \sqrt{2\gamma \eta_{\rm A}}\hat{a} - \sqrt{(1 - L_{\rm A})T}\hat{b}^{(0)}_2 - \sqrt{(1-L_{\rm A})(1-T)}\hat{b}^{(0)}_3 - \sqrt{L_{\rm A}}\hat{b}_{4}^{(0)}\,,\\
\hat{b}_{\rm B}^{(3)} &= \sqrt{2\gamma \eta_{\rm B}}\hat{a} - \sqrt{(1 - L_{\rm B})(1 - T)}\hat{b}_2^{(0)} + \sqrt{(1-L_{\rm B})T}\hat{b}_3^{(0)} - \sqrt{L_{\rm B}}\hat{b}_5^{(0)}\,.
\end{align}

With the field present at both Alice's and Bob's detectors, we can
compute the operator corresponding to their homodyne measurement. 
The procedure of homodyne measurement is briefly explained as follows. 
The input field is mixed with a strong classical field, called a 
local oscillator (LO), on a balanced beam-splitter, whereby each output is 
measured using a photodetector. The difference in the measured intensities of 
the two photodetectors gives the result of the homodyne measurement, up to a 
constant. We take the state of the strong LO to be a coherent state 
$|\beta e^{i\theta_m}\rangle$ with $\beta \in {\mathbb R}$ large enough so that the quantum fluctuations in both $\beta$ and $\theta_m$ are negligible. Here 
$\beta^2$ is the photon flux of the beam and $\theta_m$ its phase; $m = A$, 
$B$ depending on whether we are considering Alice's or Bob's measurement. 
Under this assumption, we can replace the annihilation operator describing 
the LO with its expectation value $\beta e^{i\theta_m}$. Thus, after the 
balanced beam-splitter, we have
\begin{equation}
    \begin{pmatrix}
		\hat c_m \\
		\hat d_m 
	\end{pmatrix}
	=
 \frac{1}{\sqrt{2}}
	\begin{pmatrix}
		1 & -1\\
		1 & 1 
	\end{pmatrix}
	\begin{pmatrix}
		\hat b_m^{(3)} \\
		\beta e^{i\theta_m} 
	\end{pmatrix}.
\end{equation}
The intensities of the beams at the detectors are $\hat c_m\dg \hat c_m$ and $\hat d_m\dg \hat d_m$. Thus, the operator describing Alice's and Bob's homodyne measurement is 
\begin{equation}\label{Meas_Op}
\hat{y}_m =\beta^{-1}(\hat{d}_m\dg \hat{d}_m - \hat{c}_m\dg \hat{c}_m) = e^{-i\theta_m}\hat{b}_m^{(3)} + e^{i\theta_m}(\hat{b}_m^{(3)})\dg\,.
\end{equation}

We now introduce the canonical position and momentum operators for the cavity field, defined as
\begin{equation}
    \hat{x} = \sqrt{\frac{\hbar}{2}} (\hat{a} + \hat{a}\dg)\,, \qquad \hat{p} = i\sqrt{\frac{\hbar}{2}}(\hat{a}\dg - \hat{a})\,,
\end{equation}
and for the normalized noise quadrature operators
\begin{equation}
    d\hat{x}_j^{(0)} = \left(d\hat{B}_j^{(0)} + (d\hat{B}_j^{(0)})\dg\right)\,,\qquad d\hat{p}_j^{(0)} = i\left((d\hat{B}_j^{(0)})\dg -  d\hat{B}_j^{(0)}\right)\,.
\end{equation}
Note, these noise quadrature operators are normalized such that 
\begin{align}
    \ex{d\hat{x}_j^{(0)}} = 0\,, \qquad{\rm and}\qquad \ex{(d\hat{x}_j^{(0)})^2} = dt\,,
\end{align}
where the expectation is taken with the vacuum state in the $j$-th mode, and similarly for $d\hat{p}_j^{(0)}$. With these operators, the equations for the system become
\begin{align}
d \hat{\bf x} &= A \hat{\bf x}dt + B d\hat{\bf v}\,,\label{eq:LLE}\\
\hat{y}_m dt &= C_m \hat{\bf x}dt + D_m d\hat{\bf v}\label{eq:LMC}\,,
\end{align}
where 
\begin{align}
\hat{\mathbf{x}} &\equiv \begin{pmatrix}\hat x & \hat p\end{pmatrix}\tp, \\
d\hat{\mathbf{v}} &\equiv \begin{pmatrix}d\hat{x}_1^{(0)} & d\hat{p}_1^{(0)} & \dots & d\hat{x}_5^{(0)} & d\hat{p}_5^{(0)}\end{pmatrix}\tp.
\end{align}
The matrices are
\begin{align}
&A = 
\begin{pmatrix}
	-\gamma (1-\xi) & 0\\
	0 & -\gamma (1+\xi)
\end{pmatrix},\\
&B = \sqrt{\frac{\hbar}{2}}
\begin{pmatrix}
    \sqrt{\gammaLc} & 0 & \sqrt{\gammaTc} & 0 & 0 & 0 & 0 & 0 & 0 & 0\\
     0 & \sqrt{\gammaLc} & 0 & \sqrt{\gammaTc} & 0 & 0 & 0 & 0 & 0 & 0
\end{pmatrix},\\
&C_m = \sqrt{\frac{4\gamma\eta_m}{\hbar}}
\begin{pmatrix}
	\cos\theta_m & \sin\theta_m
\end{pmatrix},\\
&D_m = 
\begin{pmatrix}
       0 &0 & D^{(m)}_{2,c}&D_{2,s}^{(m)}& D^{(m)}_{3,c}&D_{3,s}^{(m)} &D_{4,c}^{(m)} &D_{4,s}^{(m)} &D^{(m)}_{5,c}  &D_{5,s}^{(m)}
\end{pmatrix},
\end{align}
where we use the normalized pump amplitude $\xi = \epsilon/\gamma \in [0,1)$, given from the parametric gain $G_+$ as $\xi = 1-1/\sqrt{G_+}$, and 
\begin{equation}
    \begin{tabular}{ll}
        $D_{j,c}^{(m)} = d_j^{(m)}\cos\theta_m$, & $D_{j,s}^{(m)} = d_j^{(m)}\sin\theta_m$,\\
        $d_2^{(A)} = -\sqrt{(1-L_{\rm A})T}$, & $d_2^{(B)} = -\sqrt{(1-L_{\rm B})(1 - T)}$,\\
        $d_3^{(A)} = -\sqrt{(1-L_{\rm A})(1 - T)}$, &$ d_3^{(B)} = \sqrt{(1-L_{\rm B})T}$,\\
        $d_4^{(A)} = -\sqrt{L_{\rm A}}$, & $d_4^{(B)} = 0$,\\
        $d_5^{(A)} = 0$, & $d_5^{(B)} = -\sqrt{L_{\rm B}}$.
    \end{tabular}
\end{equation}

The covariances of the noise vectors (w.r.t. a vacuum state in each mode) are
\begin{align}
	\label{eq:BB}
	Q dt &= {\rm Cov}(B d \hat{\mathbf{v}},B d \hat{\mathbf{v}}) = \hbar \gamma I_2 dt, \\
	\label{eq:DD}
	R_m dt &= {\rm Cov}(D_m d \hat{\mathbf{v}},D_m d \hat{\mathbf{v}}) = dt,\\
	\label{eq:DB}
	S_m dt &= {\rm Cov}(D_m d \hat{\mathbf{v}},B d \hat{\mathbf{v}}) = -\frac{\hbar}{2}C_m dt,
\end{align}
where $I_k$ is the $k \times k$ identity matrix. Note that the noise of Alice's measurement is uncorrelated with the noise of Bob's measurement, i.e.,
\begin{equation}
    \ex{(D_{\rm A} d\hat{\bf v}) D_{\rm B} d\hat{\bf v}} = 0\,.
\end{equation}

\subsection{\label{sec:eqs}Quantum filtering and smoothing equations}
Equations \eqref{eq:LLE} and \eqref{eq:LMC} are \textit{linear} Langevin equations. This means that, provided the cavity begins in a Gaussian state, this system is a linear Gaussian quantum system\cite{wiseman2009quantum}. As such, we can utilize the linear Gaussian quantum state smoothing theory. See Ref.\cite{laverick2019quantum} for the complete derivation of the equations that are to follow. The quantum state of a continuous-variable quantum system can be represented in many different ways. In our case we are interested in the Wigner function\cite{wiseman2009quantum,walls2025quantum3e}, which for a linear Gaussian quantum system is a Gaussian distribution $W({\bf x}) = g({\bf x};\ex{\hat{\bf x}}, V)$, where the state is completely parameterized by the mean $\ex{\hat{\bf x}}$ and covariance $V$. This means that, for any estimate of the state of the system, we only need to estimate the mean and covariance matrix. 

The true mean $\xt$ and covariance $V_{\rm T}$, that is, the mean and covariance conditioned both Alice's and Bob's measurement outcomes, evolve according to the following:
\begin{align}
	\label{eq:xT_original}
	d\xt &= A\xt d t + \left(V_{\rm T} C_{\rm A}\tp+S_{\rm A}\right)R_{\rm A}^{-1} d w_{\rm A} + \left(V_{\rm T} C_{\rm B}\tp+S_{\rm B}\right)R_{\rm B}^{-1}dw_{\rm B}, \\
	\frac{d V_{\rm T}}{dt} &=AV_{\rm T} +V_{\rm T} A\tp + Q 
	-\left( V_{\rm T}C_{\rm A}\tp+S_{\rm A}\right) R_{\rm A}^{-1} \left(V_{\rm T}C_{\rm A}\tp+S_{\rm A}\right)\tp \nonumber\\
    &\hspace{15em}-\left( V_{\rm T} C_{\rm B}\tp+S_{\rm B}\right) R_{\rm B}^{-1} \left(V_{\rm T} C_{\rm B}\tp+S_{\rm B}\right)\tp.
	\label{eq:VT_original}
\end{align}
Here, Alice's innovation term $dw_{\rm A}$ is an infinitesimal Wiener increment, defined as $dw_{\rm A} = y_{\rm A} dt - C_{\rm A}\xt dt$ with $y_{\rm A}$ being the outcome Alice obtains by measuring $\hat{y}_{\rm A}$. Bob's innovation is defined similarly. The Wiener increments satisfies the following relations
\begin{equation}
\mathbb{E}[dw_m(t)] = 0\,, \qquad dw_m(t)dw_{m'}(t') = \delta_{m,m'}\delta(t-t')dt^2\,,   
\end{equation}
for $m = A,B$.
The mean $\xf$ and covariance $V_{\rm F}$ for Alice's filtered estimate are given by
\begin{align}
	\label{eq:xF}
	d\xf &= A\xf dt  + \left( V_{\rm F}C_{\rm A}\tp + S_{\rm A}\right)R_{\rm A}^{-1}d w_{\rm F}, \\
	\frac{d V_{\rm F}}{dt} &=AV_{\rm F} +V_{\rm F} A\tp + Q 
	-\left( V_{\rm F}C_{\rm A}\tp+S_{\rm A}\right)R_{\rm A}^{-1} \left(V_{\rm F}C_{\rm A}\tp+S_{\rm A}\right)\tp.
	\label{eq:VF}
\end{align}
$\xf$ and $V_{\rm F}$ are the mean and covariance of the filtered state and $dw_{\rm F} =  y_{\rm A} dt - C_{\rm A}\xf dt$.
The retrofiltering equations, using only Alice's future measurement information, are 
\begin{align}
	\label{eq:xR}
	-d\xr &= -A\xr dt  + \left( V_{\rm R} C_{\rm A}\tp - S_{\rm A}\right)R_{\rm A}^{-1}dw_{\rm R}, \\
	-\frac{d V_{\rm R}}{dt} &=-AV_{\rm R} -V_{\rm R} A\tp + Q 
	-\left( V_{\rm R} C_{\rm A}\tp - S_{\rm A}\right) R_{\rm A}^{-1} \left( V_{\rm R} C_{\rm A}\tp - S_{\rm A}\right)\tp,
	\label{eq:VR}
\end{align}
where $\xr$ and $V_{\rm R}$ is the mean and covariance of a positive operator-valued measure (POVM) element for the future record, called the retrofiltered effect. Here, the retrofiltered innovation is $dw_{\rm R} = y_{\rm A} dt - C_{\rm A}\xr dt$.

Finally, the smoothed estimate of the state, using both past and future records, is calculated via
\begin{align}
	\label{eq:xS}
	\xs &= (V_{\rm S} - V_{\rm T}) \left[ (V_{\rm F} - V_{\rm T})^{-1}\xf + (V_{\rm R}+ V_{\rm T})^{-1} \xr\right],\\
	\label{eq:VS}
	V_{\rm S} &= \left[(V_{\rm F}- V_{\rm T})^{-1} + (V_{\rm R} + V_{\rm T})^{-1}\right]^{-1} + V_{\rm T}.
\end{align}
$\xs$ and $V_{\rm S}$ are the mean and covariance of the smoothed state. Note that these equations are derived by considering the Bayesian estimation with regard to $\xt$, meaning $\xf$, $\xr$, and $\xs$ are the best estimate of $\xt$ using only past, only future, and both past and future records, respectively. The sum of the covariance of the true state $V_{\rm T}$ and the estimation error matrices $V_{\rm F} -V_{\rm T}$, $V_{\rm R} + V_{\rm T}$, and $V_{\rm S} - V_{\rm T}$, gives the covariances of the estimated state $V_{\rm F}$, $V_{\rm R}$, and $V_{\rm S}$.

\paragraph{Smoothing equations for numerical calculations}
In the previous smoothing equations, there are a couple of subtleties that need to be discussed in order to perform numerical calculations. To compute the smoothed 
state, one needs to calculate two matrix inverses, with the most problematic being $(V\fil - V\god)\inv$. This is because, especially at the initial time $t_0$, the 
matrix can be singular, i.e., $\det(V\fil - V\god) = 0$. We can resolve this issue by algebraic manipulation of Eqs.~\!\eqref{eq:xS} and \eqref{eq:VS}, yielding,
\begin{align}
	\label{eq:xS_sub1}
	\xs &=  \left[I + (V\fil - V\god)(V_{\rm R} + V_{\rm T})^{-1}\right]^{-1}\left[\xf + (V\fil - V\god)(V_{\rm R}+ V_{\rm T})^{-1} \xr\right],\\
	\label{eq:VS_sub1}
	V_{\rm S} &= \left[I + (V\fil - V\god)(V_{\rm R} + V_{\rm T})^{-1}\right]\inv (V\fil - V\god)+ V_{\rm T},
\end{align}
where $I$ is the identity matrix with the same dimension as $V$. Note, this issue is unlikely to occur in a real system with the $(V\rfil + V\god)\inv$ term, since, for this term to be singular, would require the true state to have an infinite amount of squeezing.

The second subtlety is with the final condition on the retrofiltered effect, i.e., $V\rfil(t_f) = \infty$. While the infinity is, in principle, numerically problematic, there is a deeper problem, namely, what is the final condition on the retrofiltered mean? From a theoretical perspective the final condition on the mean is meaningless as, with an infinite covariance, the Gaussian distribution becomes the uniform distribution over the real numbers. However, in a numerical computation, one would need to choose an exceedingly large but nevertheless finite value for the final covariance and the choice of $\xr$ will matter. This issue can be avoided by instead solving the new quantities
\begin{align}
	\halo{\Lambda}_{\rm R} &\equiv (V_{\rm R} + V_{\rm T})^{-1}\,, \\
	 \halo{\mathbf{z}}_{\rm R} &\equiv  \halo{\Lambda}_{\rm R} \xr\,,
\end{align}
which have final conditions $\halo{\bz}\rfil(t_f) = 0$, $\halo{\Lambda}\rfil(t_f) = 0$. Note, for consistency with the literature, we are using the notation of Refs.\cite{laverick2019quantum,laverick2021linear}.
By calculating the derivatives of $\halo{\bz}\rfil$ and $\halo{\Lambda}\rfil$ with respect to time, and using Eqs.~\!\eqref{eq:xT_original}, (\ref{eq:VT_original}), (\ref{eq:xR}) and (\ref{eq:VR}), we obtain:
\begin{align}
	-d\halo{\mathbf{z}}_{\rm R} &= \left(\bar{A}\tp - \halo{\Lambda}_{\rm R} \bar{Q}\right)\halo{\mathbf{z}}_{\rm R} dt
	+ \left[C_{\rm A}\tp - \halo{\Lambda}_{\rm R} (V_{\rm T} C_{\rm A}\tp + S_{\rm A})\right] R_{\rm A}^{-1} y_{\rm A} dt, \\
	-\frac{d \halo{\Lambda}_{\rm R}}{dt} &=\halo{\Lambda}_{\rm R}\bar{A} +\bar{A}\tp \halo{\Lambda}_{\rm R} - \halo{\Lambda}_{\rm R}\bar{Q} \halo{\Lambda}_{\rm R} + C_{\rm A}\tp R_{\rm A}^{-1} C_{\rm A}^{},
\end{align}
where 
\begin{align}
	\bar{A} &= A - \left(V_{\rm T} C_{\rm A} \tp + S_{\rm A} \right) R_{\rm A} ^{-1} C_{\rm A} , 
	\\
	\bar{Q} &= \left( V_{\rm T} C_{\rm B}\tp + S_{\rm B}\right) R_{\rm B}^{-1} \left( V_{\rm T} C_{\rm B}\tp + S_{\rm B}\right)\tp.
\end{align}
After this transformation, Eqs.~\!\eqref{eq:xS_sub1} and \eqref{eq:VS_sub1} become
\begin{align}
	\label{eq:xS2}
	\xs &= \left[I + (V_{\rm F} - V_{\rm T}) \halo{\Lambda}_{\rm R} \right]^{-1} \left[ \xf + (V_{\rm F} - V_{\rm T}) \halo{\mathbf{z}}_{\rm R}\right], \\
	V_{\rm S} &= \left[I + (V_{\rm F} - V_{\rm T}) \halo{\Lambda}_{\rm R} \right]^{-1} (V_{\rm F}- V_{\rm T}) + V_{\rm T}\\ 
	& = V\god + (V_{\rm F} - V_{\rm T}) \left[I +  \halo{\Lambda}_{\rm R} (V_{\rm F} - V_{\rm T})\right]^{-1}.
	\label{eq:VS2}
\end{align}

\subsection{Calculating the average Trace-Squared Deviation}
\subsubsection{Experimental values}
Let us consider the average Trace-Squared Deviation (TrSD) from true states defined in the main text,
\beq
{\cal D}(\rho_{\rm C}) \equiv {\mathbb E}\left[\Tr[(\rho\god - \rho_{\rm C})^2]\right] = \sum_{{\bf U}\fil,{\bf O}_{\rm C}} \wp({\bf U}\fil,{\bf O}_{\rm C}) \Tr[(\rho\god - \rho_{\rm C})^2]\,,
\eeq
where ${\mathbb E}$ denotes an ensemble average over the true states (or equivalently over all the measurement records), the subscript ${\rm C} = {\rm F,\, S}$ denotes either a filtered or smoothed conditioning, with ${\bf O}\fil$ denoting Alice's past measurement record and ${\bf O}\sm$ denoting the past-future record. Similar notation is used for Bob's record ${\bf U}$. By expanding the term in the trace, one obtains
\begin{align}
{\cal D}(\rho_{\rm C}) &= \sum_{{\bf U}\fil,{\bf O}_{\rm C}} \wp({\bf U}\fil,{\bf O}_{\rm C}) \left(\Tr[\rho\god^2] -2\Tr[\rho\god\rho_{\rm C}] + \Tr[\rho_{\rm C}^2]\right)\\
& = \sum_{{\bf U}\fil,{\bf O}_{\rm C}} \wp({\bf U}\fil,{\bf O}_{\rm C}) \left({\cal P}(\rho_{\rm T}) -2\Tr[\rho\god\rho_{\rm C}] + {\cal P}(\rho_{\rm C})\right)\label{eq:purityform}\,,
\end{align}
where the purity is defined as ${\cal P}(\rho) = \Tr[\rho^2]$. Since we are dealing with continuous-variable quantum states, it is more useful to express the quantum state in terms of its corresponding Wigner quasiprobability distribution $W(\bx)$. Using the property for a single mode system\cite{wiseman2009quantum} that $\Tr[\hat{A}\hat{B}] = 2\hbar\pi\int\dd^2\bx W_{\rm A}(\bx)W_{\rm B}(\bx)$ and assuming Gaussian Wigner distributions, we obtain
\beq
{\cal D}(\rho_{\rm C}) = \sum_{{\bf U}\fil,{\bf O}_{\rm C}} \wp({\bf U}\fil,{\bf O}_{\rm C}) \left(\frac{\hbar}{2\sqrt{|V\god|}} -4 \hbar\pi g(\ex{\hat\bx}\god;\ex{\hat\bx}_{\rm C},V\god + V_{\rm C}) + \frac{\hbar}{2\sqrt{|V_{\rm C}|}}\right)\,,
\eeq
where $g(\bx,{\bf \mu}, V)$ denotes a normalized Gaussian of a mean $\mu$ and a covariance matrix $V$ and we have used $\int \dd^2\bx g(\bx;\mu,A)g(\bx;\nu,B) = g(\mu;\nu,A+B)$. Noting that the covariance matrix does not depend on either ${\bf U}$ or ${\bf O}$, we end up with
\beq
{\cal D}(\rho_{\rm C}) =  \frac{\hbar}{2}(\sqrt{|V\god|\inv} + \sqrt{|V_{\rm C}|\inv}) -4 \hbar\pi \sum_{{\bf U}\fil,{\bf O}_{\rm C}} \wp({\bf U}\fil,{\bf O}_{\rm C}) g(\ex{\hat\bx}\god;\ex{\hat\bx}_{\rm C},V\god + V_{\rm C})\,,
\eeq
which is this expression we have used to compute the average TrSD from the experimental data. 

\subsubsection{Theoretical values}
To derive the theoretical value for the average TrSD, let us return to \eqref{eq:purityform}. Using the fact\cite{guevara2015quantum} that $\rho_{\rm C} = \sum_{{\bf U}\fil} \wp({\bf U}\fil|{\bf O}_{\rm C}) \rho\god$ and that $\wp({\bf U}\fil,{\bf O}_{\rm C}) = \wp({\bf U}\fil|{\bf O}_{\rm C})\wp({\bf O}_{\rm C})$, we can simplify \eqref{eq:purityform} to 
\begin{align}
{\cal D}(\rho_{\rm C}) &= \sum_{{\bf O}_{\rm C}} \wp({\bf O}_{\rm C}) \left( {\cal P}(\rho\god) -2\Tr[\rho_{\rm C}^2] + {\cal P}(\rho_{\rm C})\right)\\
& = \sum_{{\bf O}_{\rm C}} \wp({\bf O}_{\rm C}) \left({\cal P}(\rho\god) - {\cal P}(\rho_{\rm C})\right)\\
& = \frac{\hbar}{2}\left(\sqrt{|V\god|\inv} - \sqrt{|V_{\rm C}|\inv}\right)\,,
\end{align}
where, in the final line, we have used the fact that the covariance matrix is independent of the particular measurement record obtained.

\subsection{Squeezing of true states}
\begin{figure}[h]\centering
\includegraphics[scale=1]{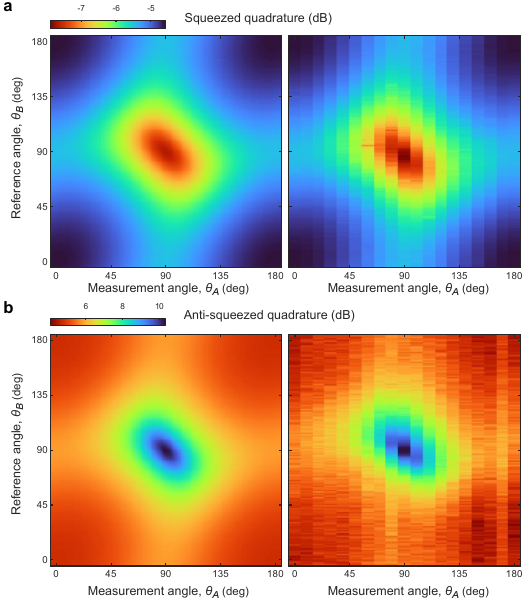}
	\caption{
    The amount of squeezing and anti-squeezing, i.e., the uncertainty in the squeezed quadrature of the true state, ${\cal S}(\rho_{\rm T})$, in \textbf{a}, and the anti-squeezed quadrature of the true state, ${\cal A}(\rho_{\rm T})$, in \textbf{b}, for different values of $\theta_{\rm A}$ and $\theta_{\rm B}$. Left is the theory prediction and right is the experimental verification.
    }
	\label{fig:truesqueeze}
\end{figure}

Here we show the squeezing of the true state, i.e., the uncertainty in the squeezed quadrature of the true states, as functions of Alice's and Bob's measurement angles in Fig.~\ref{fig:truesqueeze}. Both the theory prediction as well as the experimental verification show that the significant true-state squeezing occurs when Bob's measures a quadrature that is anti-correlated with Alice's (i.e., $\theta_{\rm B}\approx 180^\circ - \theta_{\rm A}$). 

\end{document}